\documentclass[12pt,journal]{IEEEtran}
\usepackage{epsfig,makeidx,color}
\usepackage{graphicx}
\usepackage{amsbsy}
\usepackage{amsmath}
\usepackage{amssymb}
\usepackage{euscript}
\usepackage{subfigure}
\usepackage{psfrag}

\newtheorem{mytheorem}{\bf Theorem}
\newtheorem{myresult}{\bf Result}

\newcommand {\Define} {\stackrel {\Delta} {=}  }

\begin{document}

\title{Per-antenna Constant Envelope Precoding for \\ Large
Multi-User MIMO Systems}
\author{\IEEEauthorblockN{Saif Khan Mohammed* and Erik G. Larsson}
}
\onecolumn
\maketitle
%\singlespacing
\vspace{-15mm}

\begin{abstract}
We consider the multi-user MIMO broadcast channel with $M$ single-antenna users and $N$ transmit antennas
under the constraint that each antenna emits signals having constant envelope (CE).
The motivation for this is that CE signals  facilitate the use of power-efficient RF power amplifiers.
Analytical and numerical results show that, under certain mild conditions on the channel gains,
for a fixed $M$, array gain is achievable even under
the stringent per-antenna CE constraint (essentially, for a fixed $M$, at sufficiently large $N$ the total transmitted
power can be reduced with increasing $N$ while maintaining a fixed information rate to each user).
Simulations for the i.i.d. Rayleigh fading channel show that the total transmit power can be reduced linearly
with increasing $N$ (i.e., an $O(N)$ array gain).
We also propose a precoding scheme which finds near-optimal CE signals to be transmitted, and has $O(MN)$ complexity.
Also, in terms of the total transmit power required to achieve a fixed desired information sum-rate,
despite the stringent per-antenna CE constraint, the proposed CE precoding scheme performs {\em close} to the sum-capacity
achieving scheme for an
average-only total transmit power constrained channel.
%Compared to an average only total transmit power constrained channel, the additional total transmit power
%required under the per-antenna CE constraint to achieve a desired per-user information rate is shown to be small when
%$N$ is sufficiently large.
%Specifically, for a i.i.d.~Rayleigh fading channel, with sufficiently large $N$ ($N\gtrapprox 3M$),
%the additional total transmit power 
%required under the per-antenna CE constraint is roughly $2$ dB for a wide-range of desired per-user information rate.
\end{abstract}
\vspace{-5mm}
\IEEEpeerreviewmaketitle
{\renewcommand{\thefootnote}{} \footnote{The authors are with the Communication Systems Division, Dept. of Electrical Engineering (ISY), 
Link{\"o}ping University, Link{\"o}ping, Sweden.  This work was supported by the
    Swedish Foundation for Strategic Research (SSF) and ELLIIT.
    E. G. Larsson is a Royal Swedish Academy of Sciences (KVA)
    Research Fellow supported by a grant from the Knut and Alice
    Wallenberg Foundation. Parts of the results in this paper were presented 
    at IEEE ICASSP 2012 \cite{icassp2012}.
    Also, the simpler special case of $M=1$ (i.e., single-user) has been studied by us in much greater detail in \cite{arxiv_su}.
    }}
\vspace{-15mm}

\section{Introduction}
\label{sec:intro}
We consider a Gaussian Broadcast Channel (GBC), wherein a base station (BS) having $N$ antennas communicates
with $M$ single-antenna users in the downlink.
Large antenna arrays at the BS has been of recent interest, due to their
remarkable ability to suppress multi-user interference (MUI) with very simple precoding techniques \cite{SPM-paper}.
Specifically, under an average only total transmit power constraint (APC), for a fixed $M$, a simple matched-filter precoder has been shown to achieve total MUI suppression
in the limit as $N \rightarrow \infty$ \cite{TM10}.
Additionally, due to the inherent array power gain property\footnote{\footnotesize{
Under an APC constraint, for a fixed $M$ and a fixed desired information sum-rate, the required total transmit
power decreases with increasing $N$ \cite{DTse}.}},
large antenna arrays are also being considered as an enabler for reducing power consumption
in wireless communications, especially since the operational power consumption at BS is becoming a matter of world-wide
concern \cite{GreenTouch,ComMag2}.
%\footnote{\footnotesize{Operational power consumption at BS has been reported to account for roughly ten percent of the greenhouse
%gas emissions due to ICT \cite{ComMag}}}.

Despite the benefits of large antenna arrays at BS, practically building them would
require cheap and power-efficient RF components like the power amplifier (PA).\footnote{\footnotesize{In conventional BS,
power-inefficient PA's contribute to roughly $40$-$50$ percent of the total operational power consumption \cite{ComMag2}.}}
With current technology, power-efficient RF components are generally non-linear.
The type of transmitted signal that facilitates the use of
most power-efficient/non-linear RF components, is a {\em constant envelope} (CE) signal.
In this paper, we therefore consider a GBC, where the signal transmitted from each BS antenna
has a {\em constant} amplitude for every channel-use and which is independent of the channel realization.\footnote{\footnotesize{
In this paper, we only consider the discrete-time complex baseband equivalent channel model, where
we aim to restrict the discrete-time per-antenna channel input to have no amplitude variations.
Compared to precoding methods which result in large amplitude-variations in the discrete-time channel input, the CE precoding method
proposed in this paper is expected to result in {\em continuous-time} transmit signals which have a significantly improved
peak-to-average-power-ratio (PAPR).
However, this does not necessarily mean that the proposed precoding method will result in continuous-time transmit signals having a {\em perfectly} constant envelope.
%Therefore
%appropriate smooth transmit pulse-shaping filters must be used to generate continuous-time transmit signals.
Generation of perfectly constant envelope continuous-time transmit signals has not been covered in this paper, and constitutes future work for us.
One possible method to generate almost constant-envelope continuous-time signals could be that,
in addition to constraining the discrete-time channel input to have no amplitude variations, one could also consider constraining the phase variation between consecutive symbols of the discrete-time channel input. 
}}

Since, the per-antenna CE constraint is much more restrictive than APC,
we investigate as to whether MUI suppression and array power gain can still be achieved
under the stringent per-antenna CE constraint?

To the best of our knowledge, there is no reported work which addresses this question.
Most reported work on per-antenna communication consider an average-only or a peak-only
power constraint (see \cite{WeiYu,KK} and references therein).
In this paper, firstly, we derive expressions for the MUI at each user under the per-antenna CE constraint, and then propose
a low-complexity CE precoding scheme with the objective of minimizing the MUI energy at each user.
For a given vector of information symbols to be communicated to the users, the proposed
precoding scheme chooses per-antenna CE transmit signals in such a way that the MUI energy at each user
is small.\footnote{\footnotesize{Here ``small'' implies that the MUI energy is of the same order or less than the variance of the additive white Gaussian noise (AWGN)
at the receiver. Throughout the paper, we assume that such large antenna systems will operate in a regime where
the information rate performance is not critically limited by MUI. This is because, it is highly power-inefficient to
operate in a regime where the
MUI energy is significantly more than the AWGN variance \cite{hien_paper}.}}

Secondly, under certain mild channel conditions (including i.i.d. fading), using a novel
probabilistic approach, we analytically show that,
{\em MUI suppression can be achieved even under the stringent per-antenna CE constraint.}
Specifically, for a fixed $M$ and fixed user information symbol alphabets, an arbitrarily low
MUI energy can be guaranteed at each user, by choosing a sufficiently large $N$.
Our analysis further reveals that, with a fixed $M$ and increasing $N$, the total transmitted power can be
{\em reduced}
while maintaining a constant signal-to-interference-and-noise-ratio (SINR) level at each user.

Thirdly, through simulation, we confirm our analytical observations for the i.i.d. Rayleigh fading channel.
For the proposed CE precoder, we numerically compute an achievable ergodic information sum-rate,
and observe that, for a fixed $M$ and a fixed desired ergodic sum-rate, {\em the required total transmit power reduces
linearly with increasing $N$} (i.e., achievability of an $O(N)$ array power gain under the per-antenna CE constraint).
We also observe that, to achieve a given desired ergodic information sum-rate, compared to the
optimal GBC sum-capacity achieving scheme under APC, the extra
total transmit power required by the proposed CE precoding scheme is {\em small} (roughly $2.0$ dB for sufficiently large $N$).

\textbf{Notation:}
${\mathbb C}$ and ${\mathbb R}$ denote the set of complex and real numbers.
%${\mathbb C}$ denotes the set of complex numbers.
$\vert x \vert$, $x^*$ and $\arg(x)$ denote the absolute value, complex conjugate and argument of $x \in {\mathbb C}$ respectively.
$\Vert {\bf h} \Vert^2 \Define \sum_i \vert h_i \vert^2$ denotes the squared Euclidean-norm of ${\bf h}=(h_1,\cdots,h_N) \in {\mathbb C}^{N}$.
${\mathbb E}[\cdot]$ denotes the expectation operator.
Abbreviations: r.v. (random variable), bpcu (bits-per-channel-use), p.d.f. (probability density function).
{
\vspace{4mm}
}
\section{System Model}
\label{sec:sysmodel}
{
\vspace{2mm}
}
Let the complex channel gain between the $i$-th BS antenna and the $k$-th user be denoted
by $h_{k,i}$. The vector of channel gains from the BS antennas to the $k$-th user is
denoted by ${\bf h}_k = (h_{k,1},h_{k,2},\cdots,h_{k,N})^T$.
${\bf H} \in {\mathbb C}^{M \times N}$ is the channel gain matrix with $h_{k,i}$ as its $(k,i)$-th entry.
Let $x_i$ denote the complex symbol transmitted from the $i$-th BS antenna.
Further, let $P_T$ denote the average total power transmitted from all the BS antennas.
Under APC, we must have ${\mathbb E}[\sum_{i=1}^N \vert x_i \vert^2] = P_T$,
whereas under the per-antenna CE constraint we have $\vert x_i \vert^2 = P_T/N\,,\,i=1,2,\cdots,N$ which is
clearly a more stringent constraint compared to APC.
Further, due to the per-antenna CE constraint, it is clear that $x_i$ is of the form
$x_i = \sqrt{P_T/N} e^{j \theta_i}$, where $\theta_i$ is the phase of $x_i$.\footnote{\footnotesize{
Note that CE transmission is entirely different from equal gain transmission (EGT).
We explain this difference for the simple single-user scenario ($M=1$).
In EGT a unit average energy complex information symbol $u$ is communicated to the user by transmitting $x_i = w_i \, u$ from the $i$-th
transmit antenna (with $\vert w_1 \vert = \cdots = \vert w_N \vert = \sqrt{P_T/N} $), and therefore the amplitude of the signal transmitted
from each antenna is not constant but varies with the amplitude of $u$ ($\vert x_i \vert = \sqrt{P_T/N} \vert u \vert$).
In contrast, the CE precoding method proposed in this paper (Section \ref{ce_prec_sec}) transmits a constant
amplitude signal from each antenna (i.e., $\sqrt{P_T/N} e^{j \theta_i}$ from the $i$-th antenna), where the transmit phase angles
$\theta_1,\cdots,\theta_N$ are chosen in such a way that the noise-free received signal is a known constant times the desired information symbol $u$.
}}
Under CE transmission, the symbol received
by the $k$-th user is therefore given by
\begin{eqnarray}
\label{recv_y_k}
y_k = \sqrt{\frac{P_T}{N}} \sum_{i=1}^N h_{k,i} e^{j \theta_i} + w_k \,\,,\,\,k=1,2,\ldots,M
\end{eqnarray}
where $w_k \sim {\mathcal C}{\mathcal N}(0,\sigma^2)$ is the AWGN noise at the $k$-th receiver.
For the sake of notation, let $\Theta = (\theta_1,\cdots,\theta_N)^T$ denote the vector of transmitted phase angles.
Let ${\bf u}= (\sqrt{E_1}u_1,\cdots,\sqrt{E_M}u_M)^T$ be the vector of scaled information symbols, with $u_k \in {\mathcal U}_k$ denoting the information symbol
to be communicated to the $k$-th user. Here ${\mathcal U}_k$ denotes the unit average energy information alphabet of the $k$-th user.
$E_k,k=1,2,\ldots,M$ denotes the information symbol energy for each user.
Also, let ${\mathcal U} \Define \sqrt{E_1}{\mathcal U}_1 \times \sqrt{E_2}{\mathcal U}_2 \times \cdots \times \sqrt{E_M}{\mathcal U}_M$.
Subsequently, in this paper, we are interested in scenarios where $M$ is fixed and $N$
is allowed to increase.
Also, throughout this paper, for a fixed $M$, the alphabets ${\mathcal U}_1,\cdots,{\mathcal U}_M$ are also fixed
and do not change with increasing $N$.
%\section{Proposed CE Precoding Scheme}
%\label{ce_prec_sec}
%If we adaptively choose ${\mathcal U} \subset {\mathcal M}({\bf H})$ then it is clear that
%for any information symbol vector ${\bf u} \in {\mathcal U}$
%there exists a vector of transmit phase angles $\Theta^u = (\theta_1^u,\theta_2^u,\cdots,\theta_N^u)^T$, such that
%\begin{eqnarray}
%\sqrt{E_k} u_k = \frac{  \sum_{i=1}^N h_{k,i} e^{j \theta_i^u}} {\sqrt{N}}\,\,,\,\, k=1,2,\ldots,M
%\end{eqnarray}
%and therefore, the received noise-free signal at the $k$-th user ($k=1,2,\ldots,M$) i.e.,
%is $\sqrt{P_T} \sqrt{E_k} u_k$, i.e., there is no multi-user interference from information symbols intended
%for the other users.
%
%However, due to difficulty in characterizing the region ${\mathcal M}({\bf H})$ in closed-form,
%subsequently in this paper we assume that, for a given $M$, the alphabet sets ${\mathcal U}_k \,,\,k=1,2,\ldots,M$ are fixed,
%and donot change from one channel realization to another. The alphabet sets also donot change with increasing $N$.
%Also, for a given $M$ and $N$, $E_k$ does not change from one channel realization to another.
%However, for a given $M$ and $N$, based upon performance
%measures like the ergodic information sum rate, the optimal $E_k\,,\,k=1,2,\cdots,M$ is computed a-priori (we shall see this later).
%
%\vspace{-3mm}
{
\vspace{4mm}
}
\section{MUI Analysis and the Proposed CE Precoder}
\label{muianalysis}
{
\vspace{2mm}
}
For any given information symbol vector ${\bf u}$ to be communicated,
with $\Theta$ as the transmitted phase angle vector, using (\ref{recv_y_k}) the received signal at the $k$-th user can be expressed as
%\vspace{-2mm}
\begin{eqnarray}
\label{yk_model}
y_k  =  \sqrt{P_T} \sqrt{E_k} u_k + \sqrt{P_T} s_k + w_k \,\,,\,\,
s_k  \Define   {\Big (} \frac{ \sum_{i=1}^N h_{k,i} e^{j \theta_i}  } { \sqrt{N}} - \sqrt{E_k} u_k   {\Big )}
\end{eqnarray}
where $\sqrt{P_T} \, s_k$ is the MUI term at the $k$-th user.
In this section, for any general CE precoding scheme where the signal transmitted from each BS antenna has constant envelope,
through analysis, we aim to get a better understanding of the MUI energy level at each user. Towards this end, we firstly
study the range of values taken by the noise-free received signal
at the users (scaled down by $\sqrt{P_T}$). This range of values is given by the set
\begin{eqnarray}
\label{MH_def}
{\mathcal M}({\bf H})  & \Define &  {\Big \{} {\bf v}=(v_1,\cdots,v_M) \in {\mathbb C}^M \,\,{\big |}\,\, \,
{ v_k} = \frac {\sum_{i=1}^N h_{k,i} e^{j \theta_i}}{\sqrt{N}} \,\,,\,\theta_i \in [-\pi,\pi) \,,\,i=1,\ldots,N{\Big \}} 
%{\bf e}(\Theta) & \Define & (e^{j \theta_1}, e^{j \theta_2},\cdots, e^{j \theta_N})^T.
\end{eqnarray}
For any vector ${\bf v} = (v_1,v_2,\cdots, v_M)^T \in {\mathcal M}({\bf H})$, from (\ref{MH_def})
it follows that there exists a $\Theta^v= (\theta_1^v, \cdots, \theta_N^v)^T$ such that 
%\begin{equation}
%\label{eq_v_k}
${ v_k} = \frac {\sum_{i=1}^N h_{k,i} e^{j \theta_i^v}}{\sqrt{N}} \,\,,\,\, k=1,2,\ldots,M$.
%\end{equation}
This sum can now be expressed as a sum of
$N/M$ terms (without loss of generality let us assume that $N/M$ is integral only for the argument
presented here)
\begin{eqnarray}
\label{eq_sum}
v_k  =  \sum_{q=1}^{N/M} v_k^q \,\,\,,\,\,\,
v_k^q \Define {\Big (} \hspace{-6mm}  \sum_{r=(q-1)M+1}^{qM} \hspace{-5mm} h_{k,r} e^{j \theta_r^v} \, {\Big )} \, / \, \sqrt{N} \,,\,q=1,\ldots,\frac{N}{M}.
\end{eqnarray}
From (\ref{eq_sum}) it immediately follows that ${\mathcal M}({\bf H})$ can be expressed
as a direct-sum of $N/M$ sets, i.e.
\begin{eqnarray}
\label{direct_sum_eq}
{\mathcal M}({\bf H}) & = & {\mathcal M}{\big (}{\bf H}^{(1)}{\big )} \oplus {\mathcal M}{\big (}{\bf H}^{(2)}{\big )} \oplus \cdots \oplus {\mathcal M}{\big (}{\bf H}^{(N/M)}{\big )} \nonumber \\
& & \nonumber \\
{\mathcal M}{\big (}{\bf H}^{(q)}{\big )} & \hspace{-3mm} \Define & \hspace{-4mm} {\Big \{} {\bf v}=(v_1,\cdots,v_M) \in {\mathbb C}^M \,\,{\big |}\,\, \, \,
 { v_k} = \frac {\sum_{i=1}^M h_{k,(q-1)M+i} \hspace{1mm} e^{j \theta_i}}{\sqrt{N}} \,,\,\theta_i \in [-\pi,\pi) {\Big \}} \,,\,q=1,\ldots,N/M \nonumber \\
\end{eqnarray}
where ${\bf H}^{(q)}$ is the sub-matrix of ${\bf H}$ containing only the columns numbered $(q -1)M + 1,(q -1)M + 2, \cdots, qM$.
${\mathcal M}{\big (}{\bf H}^{(q)}{\big )} \subset {\mathbb C}^M$ is the dynamic range of the received noise-free signals
when only the $M$ BS antennas numbered $(q-1)M+1,(q-1)M+2,\cdots,qM$ are used and the remaining $N - M$ antennas are inactive.
If the statistical distribution of the channel gain vector from a BS antenna to all the users is identical for all the BS antennas (as in i.i.d. channels),
then, on an average the sets ${\mathcal M}{\big (}{\bf H}^{(q)}{\big )} \,,\,q=1,\ldots,N/M$ would all have similar topological properties.
Since, ${\mathcal M}({\bf H})$ is a direct-sum of $N/M$ topologically similar sets, it is expected that for a fixed $M$, on an average the
region ${\mathcal M}({\bf H})$ expands with increasing $N$.
Specifically, for a fixed $M$ and increasing $N$, the maximum Euclidean length of any vector in ${\mathcal M}({\bf H})$ grows as $O(\sqrt{N})$, since
${\mathcal M}({\bf H})$ is a direct-sum of $O(N)$ topologically similar sets (${\mathcal M}({\bf H}^{(q)}) \,,\,q=1,2,\ldots,N/M$) with the
maximum Euclidean length of any vector in ${\mathcal M}{\big (}{\bf H}^{(q)}{\big )}$ being $O(1/\sqrt{N})$ (note that in the definition of ${\mathcal M}{\big (}{\bf H}^{(q)}{\big )}$
in (\ref{direct_sum_eq}), each component of any vector ${\bf v} \in {\mathcal M}{\big (}{\bf H}^{(q)}{\big )}$
is scaled down by $\sqrt{N}$).
Also, for a fixed $M$ and increasing $N$, since ${\mathcal M}({\bf H})$ is a direct-sum of $N/M$ similar sets,
it is expected that the set ${\mathcal M}({\bf H})$ becomes increasingly dense (i.e., the number
of elements of ${\mathcal M}({\bf H})$ in a fixed volume in ${\mathbb C}^M$ is expected to increase with increasing $N$).
The above discussion leads us to the following results in Section \ref{remark1_sec} and \ref{remark2_sec}.
{
%\vspace{-4mm}
}
{
\vspace{2mm}
}
\subsection{Diminishing MUI with increasing $N$, for fixed $M$ and fixed $E_k (k=1,\ldots,M)$}
\label{remark1_sec}
{
\vspace{2mm}
}
For a fixed $M$ and fixed $E_k$, the information alphabets and the information symbol energies are fixed.
However, since increasing $N$ (with fixed $M$) is expected to enlarge the set ${\mathcal M}({\bf H})$ and make it increasingly denser, it is highly probable
that at sufficiently large $N$, for any fixed information symbol vector ${\bf u} = (\sqrt{E_1} u_1, \cdots,\sqrt{E_M} u_M)^T \in {\mathcal U}$
there exists a vector ${\bf v} \in {\mathcal M}({\bf H})$ such that ${\bf v}$ is very {\em close} to ${\bf u}$ in terms of Euclidean distance. 
%Hence it is expected that, with increasing $N$ the MUI energy at each user must also decrease.
This then implies that, with increasing $N$ and fixed $M$, for any ${\bf u} \in {\mathcal U}$ there exists a transmit phase angle vector
$\Theta$ such that the sum of the MUI energy for all users is small compared to the AWGN variance at the receiver.
Hence, for a fixed $M$ and fixed $E_k$, it is expected that the MUI energy for each user decreases with increasing $N$.

This is in fact true, as we prove it formally for channels satisfying the following mild conditions.
Specifically for a fixed $M$, we consider a sequence of channel gain matrices $\{ {\bf H}_N \}_{N=M}^{\infty} $ satisfying
\begin{eqnarray}
\label{assumptions}
\lim_{N \rightarrow \infty} \frac{\vert {{\bf h}_{k}^{(N)}}^H {\bf h}_{l}^{(N)}\vert } {N }  & = & 0 \,\,,\,\,\forall \, k \ne l \,\,,\,\, k,l \, \in \, (1,\ldots,M) \,\,\,\, \mbox{(As.1)} \nonumber \\
\lim_{N \rightarrow \infty} \frac {\sum_{i=1}^N  \vert h_{{k_1},i}^{(N)} \vert \,  \vert h_{{l_1},i}^{(N)} \vert \,  \vert h_{{k_2},i}^{(N)} \vert \, \vert h_{{l_2},i}^{(N)} \vert }
 {  N^2 }  &  = &  0 
\,\,,\,\,\forall k_1,l_1,k_2,l_2 \in (1,2,\ldots,M) \,\, \mbox{(As.2)} \nonumber \\
\lim_{N \rightarrow \infty} \frac{\Vert {\bf h}_{k}^{(N)}  \Vert^2  } {N } &  = &  c_k \,\,,\,\, k=1,2,\ldots,M \,\,  \mbox{(As.3)}
\end{eqnarray}
where $c_k$ are positive constants, ${\bf h}_{k}^{(N)}$ denotes the $k$-th row of ${\bf H}_N$ and $h_{{k},i}^{(N)}$ denotes the $i$-th component of ${\bf h}_{k}^{(N)}$.
From the law of large numbers, it follows that i.i.d. channels satisfy these conditions with probability one \cite{Patrick}.
Physical measurements of the channel characteristics with large antenna arrays at the BS have revealed closeness to the i.i.d.
fading model, as long as the BS antennas are sufficiently spaced apart (usually half of the carrier wavelength) \cite{Fredrik,SPM-paper}.
\begin{mytheorem}\label{main_theorem1}
For a fixed $M$ and increasing $N$, consider a sequence of channel gain matrices $\{ {\bf H}_N \}_{N=M}^{\infty} $ satisfying the mild conditions
in (\ref{assumptions}).
For any given fixed finite alphabet ${\mathcal U}$ (fixed $E_k,k=1,\ldots,M$) and any given $\Delta > 0$, there exist a corresponding integer $N(\{ {\bf H}_{N} \}, {\mathcal U}, \Delta)$
such that with $N \geq N(\{ {\bf H}_{N} \}, {\mathcal U}, \Delta)$ and ${\bf H}_N$ as the channel gain matrix, for any ${\bf u} \in {\mathcal U}$ to be communicated,
there exist a phase angle vector
$\Theta_N^u(\Delta)$ = $(\theta_1^u(\Delta),\cdots,\theta_N^u(\Delta))^T$ which when transmitted,
results in the MUI energy at each user being upper bounded by $2 \Delta^2$, i.e.
%{\vspace{-1mm} }
\begin{eqnarray}
\label{thm1_eqn}
{\Big \vert } \frac{ \sum_{i=1}^N h_{k,i}^{(N)} e^{j \theta_i^u(\Delta)}}   { \sqrt{N}} - \sqrt{E_k} u_k   {\Big \vert }^2 \leq 2 \Delta^2 \,\,,\,\,k=1,\ldots,M.
\end{eqnarray}
%where $h_{k,i}^{(N)}$ denotes the $i$-th component of ${\bf h}_{k}^{(N)}$.
\end{mytheorem}

{\it Proof} --
The proof relies on technical results
stated and proved in Appendix \ref{app_conv_zN} and \ref{box_event_proof}.
All these results assume a fixed $M$ (number of user terminals) and increasing $N$ (number of BS antennas).
These results are stated for a fixed sequence of channel matrices $\{ {\bf H}_N \}_{N=M}^{\infty}$,
fixed information alphabets ${\mathcal U}_1,\cdots,{\mathcal U}_M$ and fixed information symbol energy
$E_1,\cdots,E_M$.
Further, the sequence of channel matrices $\{ {\bf H}_N \}_{N=M}^{\infty}$ is assumed to satisfy the
conditions in (\ref{assumptions}) and the information alphabets are assumed to be finite/discrete.
The proofs use a novel probabilistic approach, treating the transmitted phase angles as random variables.
We now present the proof of Theorem \ref{main_theorem1}.

Let us consider a probability space
with the transmitted phase angles $\theta_i,i=1,2,\ldots,N$ being i.i.d.
r.v's uniformly distributed in $[-\pi \,,\, \pi)$.
For a given sequence of channel matrices $\{ {\bf H}_N \}$, we define a corresponding sequence of r.v's $\{ {\bf z}_N \}$, with
${\bf z}_N \Define (z_{1}^{I^{(N)}},z_{1}^{Q^{(N)}},\ldots,z_{M}^{I^{(N)}},z_{M}^{Q^{(N)}}) \in {\mathbb R}^{2M}$, where we have
{
\vspace{2mm}
\small
\begin{eqnarray}
\label{zN_def}
z_{k}^{I^{(N)}} & \Define &  \mbox{Re}{\Big (} \frac{\sum_{i=1}^N h_{k,i}^{(N)} e^{j \theta_i}} {\sqrt{N}}  {\Big )} \,\,,\,\,
z_{k}^{Q^{(N)}}  \Define  \mbox{Im}{\Big (} \frac{\sum_{i=1}^N h_{k,i}^{(N)} e^{j \theta_i}} {\sqrt{N}}  {\Big )}, k=1,\ldots,M.
\end{eqnarray}
%\vspace{2mm}
}
\normalsize
From Theorem \ref{thm_conv_zN} in Appendix \ref{app_conv_zN} it follows that, for any channel sequence $\{ {\bf H}_N \}$ satisfying the
conditions in (\ref{assumptions}), as $N \rightarrow \infty$ (with fixed $M$), the corresponding sequence of r.v's $\{ {\bf z}_N \}$
converges
in distribution to a $2M$-dimensional real Gaussian random vector $X = (X_1^I,X_1^Q,\cdots,X_M^I,X_M^Q)^T$ with independent zero-mean
components and $\mbox{var}(X_k^I) = \mbox{var}(X_k^Q) = c_k/2 \,,\,k=1,2,\ldots,M$.
%Since the range space of the $2M$-dimensional real Gaussian random vector $X$ is ${\mathbb R}^{2M}$ and
%${\bf z}_N$ converges to $X$ in distribution, it follows that the limiting range space (as $N \rightarrow \infty$) of
%${\bf z}_N$ is also ${\mathbb R}^{2M}$.
%Further, for any
%value taken by ${\bf z}_N$, say ${\bf z}_N={\bf e}=(e_1^I,e_1^Q,\cdots,e_M^I,e_M^Q)^T$, from the definition of ${\mathcal M}({\bf H})$ in (\ref{MH_def}), it is clear that
%the complex vector ${\bf e}^c = (e_1^c,e_2^c,\cdots,e_M^c)^T$ ($e_k^c \Define (e_k^I + j e_k^Q)$) belongs to ${\mathcal M}({\bf H}_N)$.
%Hence, it can be concluded that ${\mathcal M}({\bf H}_N) \rightarrow {\mathbb C}^M$ as $N \rightarrow \infty$ (supports our observation
%in Section \ref{muianalysis} on the enlargement of ${\mathcal M}({\bf H})$ with increasing $N$).
For a given ${\bf u}=(\sqrt{E_1}u_1,\cdots,\sqrt{E_M}u_M)^T \in {\mathcal U}$, and $\Delta > 0$, we next consider the box
{
\vspace{2mm}
\small
\begin{eqnarray}
\label{box_eqn}
{\mathcal B}_{_{\Delta}}({\bf u})  \Define  {\Bigg \{} {\bf b}=(b_1^I,b_1^Q,\cdots,b_M^I,b_M^Q)^T \in {\mathbb R}^{2M} \, {\Big |}\, 
\,\, \vert b_k^I - \sqrt{E_k} u_k^I \vert \leq \Delta \,\,,\,\, \vert b_k^Q - \sqrt{E_k} u_k^Q \vert \leq \Delta 
\,\,,\,\,k=1,2,\ldots,M     {\Bigg \}} 
%& & \mbox{where} \,\,\,\, u_k^I \Define \mbox{Re}(u_k) \,,\, u_k^Q \Define \mbox{Im}(u_k) 
\end{eqnarray}
%\vspace{2mm}
}
\normalsize
where $u_k^I \Define \mbox{Re}(u_k) \,,\, u_k^Q \Define \mbox{Im}(u_k)$. The box ${\mathcal B}_{_{\Delta}}({\bf u})$
contains all those vectors in ${\mathbb R}^{2M}$ whose component-wise displacement from ${\bf u}$ is upper bounded by $\Delta$.
Using the fact that ${\bf z}_N$ converges in distribution to a Gaussian r.v. with ${\mathbb R}^{2M}$ as its range space,
in Theorem \ref{thm_continuity} (Appendix \ref{box_event_proof}) it is shown that,
for any $\Delta > 0$, there exist an integer $ N(\{ {\bf H}_N \},{\mathcal U}, \Delta)$, such that
for all $N \geq N(\{ {\bf H}_N \},{\mathcal U}, \Delta)$
{
\vspace{2mm}
\small
\begin{eqnarray}
\label{pos_prob_eq}
\mbox{Prob}({\bf z}_N \in {\mathcal B}_{_{\Delta}}({\bf u})) > 0 \,\,,\,\, \forall \, {\bf u} \in {\mathcal U}.
\end{eqnarray}
%\vspace{2mm}
}
\normalsize
Since the probability that ${\bf z}_N$ lies in the box ${\mathcal B}_{_{\Delta}}({\bf u})$ is {\em strictly positive} for all ${\bf u} \in {\mathcal U}$,
from the definitions of ${\mathcal B}_{_{\Delta}}({\bf u}) $ in (\ref{box_eqn}) and ${\bf z}_N$ in (\ref{zN_def})
it follows that, for any ${\bf u} \in {\mathcal U}$ there exist a phase angle vector $\Theta_N^u(\Delta)=(\theta_1^u(\Delta),\cdots,\theta_N^u(\Delta))^T$
such that
{
\vspace{2mm}
\small
\begin{eqnarray}
{\Big \vert} \mbox{Re}{\Big (} \frac{\sum_{i=1}^N h_{k,i}^{(N)} e^{j \theta_i^u(\Delta)}} {\sqrt{N}}  {\Big )} - \sqrt{E_k} u_k^I {\Big \vert}   \leq  \Delta
\,\,\,,\,\,\,
{\Big \vert} \mbox{Im}{\Big (} \frac{\sum_{i=1}^N h_{k,i}^{(N)} e^{j \theta_i^u(\Delta)}} {\sqrt{N}}  {\Big )} - \sqrt{E_k} u_k^Q {\Big \vert}   \leq  \Delta
\end{eqnarray}
\vspace{2mm}
}
\normalsize
for all $k=1,2,\cdots,M$, which then implies (\ref{thm1_eqn}).
$\hfill\blacksquare$

Since Theorem \ref{main_theorem1} is valid for any $\Delta > 0$
and (\ref{thm1_eqn}) holds for all $N \geq N(\{ {\bf H}_{N} \}, {\mathcal U}, \Delta)$,
we can satisfy (\ref{thm1_eqn}) for any arbitrarily small $\Delta > 0$ by having $N \geq N(\{ {\bf H}_{N} \}, {\mathcal U}, \Delta)$ i.e., a sufficiently large $N$.
Hence,
the {\em MUI energy at each user can be guaranteed to be arbitrarily small by having a sufficiently large $N$}.
Theorem 1 therefore motivates us to propose precoding techniques which can achieve small MUI energy levels as guaranteed
by the theorem.
{
\vspace{2mm}
}
\subsection{Proposed CE Precoding Scheme}
\label{ce_prec_sec}
{
\vspace{2mm}
}
For reliable communication to each user, the precoder at the BS must choose a $\Theta$ such that the MUI energy $\vert s_k \vert$ is as small as possible
for each $k=1,2,\ldots,M$.
This motivates us to consider the following non-linear least squares (NLS) problem, which for a given ${\bf u}$ to be communicated,
finds the transmit phase angles that minimize the sum of the MUI energy for all users:
%\vspace{-3mm}
\begin{eqnarray}
\label{NLS}
\Theta^{u} & = & (\theta_1^{u},\cdots,\theta_N^{u})  =  \arg \min_{\theta_i \in [-\pi,\pi) \, , \, i=1,\ldots,N} g(\Theta,{\bf u}) \nonumber \\
g(\Theta,{\bf u}) & \Define & \sum_{k=1}^M   {\Big \vert}  s_k {\Big \vert}^2 \,\, = \,\, \sum_{k=1}^M {\Big \vert} \frac{ \sum_{i=1}^N h_{k,i} e^{j \theta_i}  } { \sqrt{N}} - \sqrt{E_k} u_k {\Big \vert}^2.
\end{eqnarray}
This NLS problem is non-convex and has multiple local minima.
However, as the ratio $N/M$ becomes large, due to the large number of extra degrees of freedom ($N - M$),
the value of the objective function $g(\Theta,{\bf u})$ at most local minima has been observed to be small,
enabling gradient descent based methods to be used.\footnote{\footnotesize{
This observation is expected, since the strict positivity of the box event probability in (\ref{pos_prob_eq}) (proof of Theorem 1),
implies that there are many distinct transmit phase angles $\Theta$ such that the received noise-free vector lies
in a small $2M$-dimensional cube (box) centered at the desired information symbol vector ${\bf u}$, i.e., the MUI energy at each
user is small for many different $\Theta$.}}
However, due to the slow convergence of gradient descent based methods, we propose a novel iterative method,
which has been experimentally
observed to achieve similar performance as the gradient descent based methods, but with a significantly faster convergence.

In the proposed iterative method to solve (\ref{NLS}), we start with the $p=0$-th iteration,
where we initialize all the angles to $0$.
Each iteration consists of $N$ sub-iterations. Let $\Theta^{(p,q)} = (\theta_1^{(p,q)},\cdots,\theta_N^{(p,q)})^T$
denote the phase angle vector
after the $q$-th sub-iteration ($q=1,2,\ldots,N$) of the $p$-th iteration (subsequently we shall refer to the
$q$-th sub-iteration of the $p$-th iteration as the $(p,q)$-th iteration).
After the $(p,q)$-th iteration, the algorithm moves either to the $(p,q+1)$-th iteration (if $q < N $), or else
it moves to the $(p+1,1)$-th iteration.
In general, in the $(p,q+1)$-th iteration, the algorithm attempts to reduce the current value of the objective function
i.e., ${g}(\Theta^{(p,q)},{{\bf u}})$ by only modifying the $(q+1)$-th phase angle (i.e., $\theta_{q+1}^{(p,q)}$) while keeping the other phase angles
fixed to their values from the previous iteration. The new phase angles after the $(p,q+1)$-th iteration,
are therefore given by
\begin{eqnarray}
\label{phase_ang_update}
\hspace{2mm} \theta_{q+1}^{(p,q+1)}  & = & \mbox{arg min}_{_{_{_{\hspace{-12mm}\Theta={\big (}\theta_1^{(p,q)},\cdots,\theta_q^{(p,q)},\phi,\theta_{q+2}^{(p,q)},\cdots,\theta_N^{(p,q)}{\big )}^T \, ,\, \phi \in [-\pi, \pi)}}}} \hspace{-40mm}  {g}(\Theta,{ {\bf u}}) \nonumber \\
& = & \pi  +  \arg{\Bigg (} \sum_{k=1}^M \frac {h_{k,q+1}^*}{\sqrt{N}}  {\Big [} \, {\Big (} \,\frac{1}{\sqrt{N} }   { \mathop {\sum} \limits_{{i=1,\ne (q+1)}}^N  } h_{k,i} \, e^{j \theta_i^{(p,q)}} \, {\Big )} \, - \sqrt{E_k} {u_k}   {\Big ]}   {\Bigg )} \nonumber \\
\hspace{2mm} \theta_{i}^{(p,q+1)}  & =  & \theta_{i}^{(p,q)} \,\,,\,\,i=1,2,\ldots,N\,,\, i \ne q+1.
\end{eqnarray}
%The $q+1$-th phase angle is therefore updated as
%{
%\small
%\begin{eqnarray*}
%\theta_{q+1}^{(p,q+1)} = \pi + \arg{\Bigg (} \sum_{k=1}^M \frac {h_{k,q+1}^*}{\sqrt{N}}  {\Big [} \frac{\sum_{_{_{{i=1,\ne (q+1)}}}}^N \hspace{-10mm}  h_{k,i} e^{j \theta_i^{(p,q)}}} {\sqrt{N} } - \sqrt{E_k} {u_k}   {\Big ]}   {\Bigg )}.
%\end{eqnarray*}
%}
The algorithm is terminated after a pre-defined number of iterations.\footnote{\footnotesize{
Experimentally, we have observed that, for the i.i.d. Rayleigh fading channel, with a sufficiently large $N/M$ ratio,
beyond the $p=L$-th iteration (where $L$ is some constant integer),
the incremental reduction in the value of the objective function is minimal.
Therefore, we terminate at the $L $-th iteration. Since there are totally $LN$ sub-iterations,
from the phase angle update equation in (\ref{phase_ang_update}), it follows that the complexity of the proposed iterative algorithm is $O(MN)$.}}
We denote the phase angle vector after the last iteration by
${\widehat \Theta^{u}} = ({\widehat \theta_1^{u}},\cdots,{\widehat \theta_N^{u}} )^T$.

With ${\widehat \Theta^{u}}$ as the transmitted phase angle vector,
the received signal and the MUI term are given by
\begin{eqnarray}
\label{yk_model_hat}
y_k  =  \sqrt{P_T} \sqrt{E_k} u_k + \sqrt{P_T} {\widehat s_k} + w_k \,\,,\,\,
{\widehat s_k}  \Define   {\Big (} \frac{ \sum_{i=1}^N h_{k,i} e^{j {\widehat \theta_i^{u}} }  } { \sqrt{N}} - \sqrt{E_k} u_k   {\Big )}
\end{eqnarray}
The received signal-to-noise-and-interference-ratio (SINR) at the $k$-th user is therefore given by
{
%\vspace{-2mm}
}
\begin{eqnarray}
\label{sinr_eq}
\gamma_k({\bf H},E,\frac{P_T}{\sigma^2}) & = &\frac { E_k} { {\mathbb E}_{_{u_1,\cdots,u_M}}{\big [} {\vert}  {\widehat s_k} { \vert}^2 {\big ]}  + \frac{\sigma^2}{P_T} } 
\end{eqnarray}
where $E \Define (E_1,E_2,\cdots,E_M)^T$ is the vector of information symbol energies.
Note that the above SINR expression is for a given channel realization ${\bf H}$.
%From (\ref{sinr_eq}), it is clear that, for a fixed $P_T/\sigma^2$, the average SINR at the $k$-th user is dependent upon the mean energy of the
%multi-user interference term ${\widehat s_k}$, as well as $E_k$.
For each user, we would be ideally interested to have a low value of the MUI energy ${\mathbb E}[ \vert {\widehat s_k} \vert^2]$, since this would
imply a larger SINR.
%Based on (\ref{sinr_eq}), we have the following two important remarks which are formally shown to be true in Sections \ref{dim_sec} and \ref{inc_sec}.

\begin{figure}[t]
\begin{center}
\epsfig{file=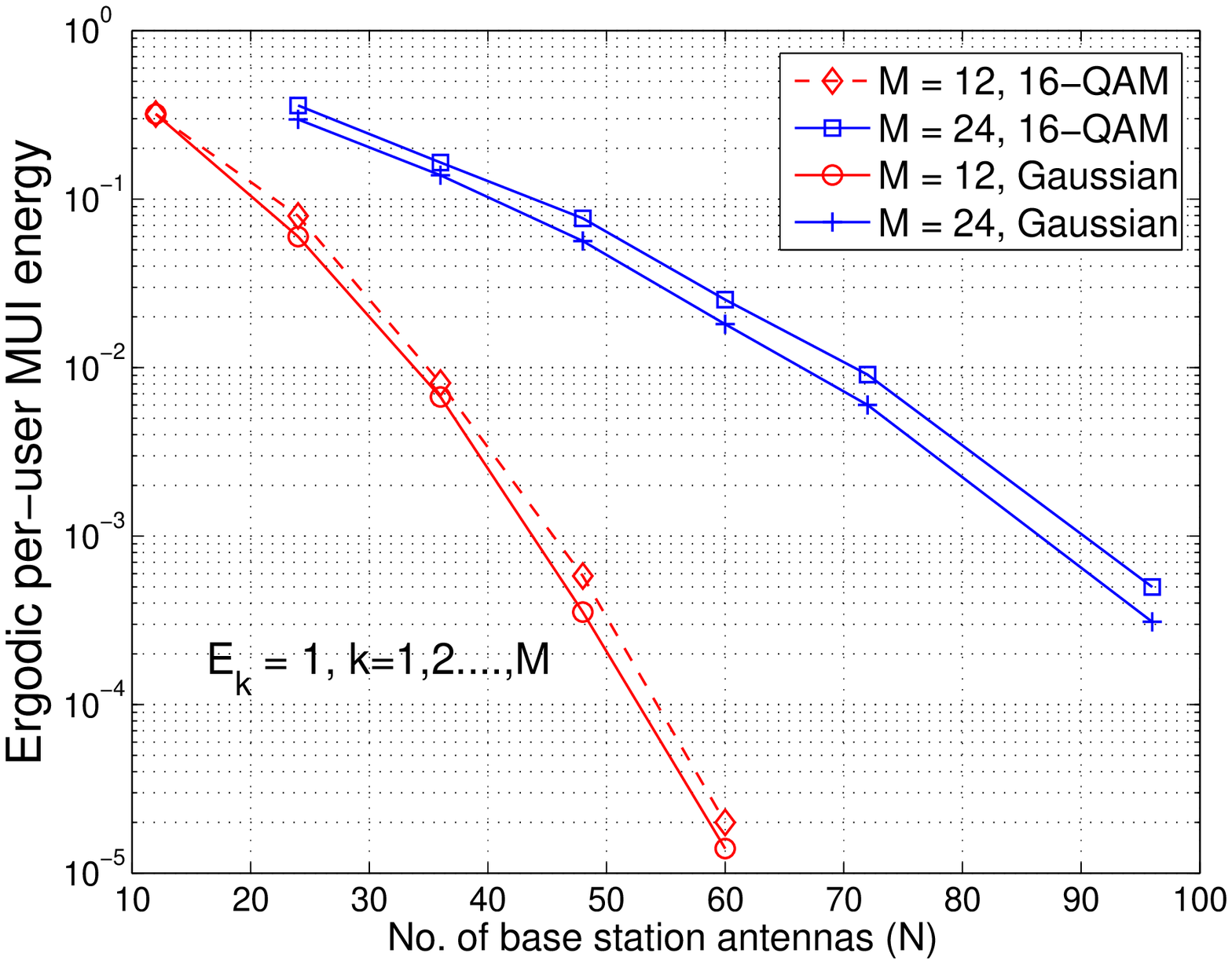, width=155mm,height=120mm}
\end{center}
\vspace{-6mm}
\caption{Reduction in the ergodic per-user MUI energy ${\mathbb E}_{{\bf H}}{\big [} {\vert}  {\widehat s_k} { \vert}^2 {\big ]}$ with increasing $N$. Fixed $M$, fixed ${\mathcal U}_1=\cdots={\mathcal U}_M=\mbox{$16$-QAM},\mbox{Gaussian}$ and fixed $E_k = 1 \,,\,k=1,2,\ldots,M$. IID ${\mathcal C}{\mathcal N}(0,1)$ Rayleigh fading.}
\label{fig_0}
\vspace{-4mm}
\end{figure}

To illustrate the result of Theorem 1, in Fig.~\ref{fig_0}, for the i.i.d. ${\mathcal C}{\mathcal N}(0,1)$ Rayleigh fading channel,
with fixed information alphabets
${\mathcal U}_1 = {\mathcal U}_2 = \cdots = {\mathcal U}_M = (\mbox{16-QAM and Gaussian})$ and
fixed information symbol energy $E_k=1,k=1,\ldots,M$,
we plot the ergodic (averaged over channel statistics) MUI energy ${\mathbb E}_{\bf H}[\vert {\widehat s_k} \vert^2]$
with the proposed CE precoding scheme (using the discussed iterative method for solving (\ref{NLS}))
as a function of increasing $N$ (${\widehat s_k}$ is given by (\ref{yk_model_hat})).\footnote{\footnotesize{ We have observed
that ${\mathbb E}_{\bf H}[\vert {\widehat s_k} \vert^2]$ is the same for all $k=1,\ldots,M$.}}
It is observed that, for a fixed $M$, fixed information alphabets and fixed information symbol energy, the ergodic per-user MUI energy decreases with increasing number
of BS antennas $N$.
This is observed to be true, not only for a finite/discrete $16$-QAM information symbol alphabet, but also for the non-discrete
Gaussian information alphabet.
%Since the MUI energy is plotted on a log scale, the linear slope of the ergodic MUI energy curves suggest that the ergodic MUI energy decreases
%exponentially in $N$.
{
\vspace{2mm}
}
\subsection{Increasing $E_k$ with increasing $N$, for a fixed $M$, fixed ${\mathcal U}_1,\cdots,{\mathcal U}_M$ and fixed desired MUI energy level}
\label{remark2_sec}
{
\vspace{2mm}
}
%In (\ref{sinr_eq}), with increase in $E_k$, the numerator of the SINR expression increases, but at the same time
%multi-user interference energy term in the denominator may also increase if $E_k$ is so large that
%$(\sqrt{E_1}u_1,\cdots,\sqrt{E_M}u_M)^T$ lies outside the region ${\mathcal M}({\bf H})$.
%There is therefore a tradeoff between the information symbol energy and the energy of the multi-user interference term.
%This then suggests that there exists an optimal value of $E_k\,,\,k=1,2,\cdots,M$ with respect to maximizing the
%average SINR for all the users.
%From Remark 1, we know that with increasing $N$ but fixed $E_k$ ($k=1,2,\cdots,M$), the multi-user interference energy decreases with high probability.
%Therefore, it can be argued that, since the set ${\mathcal M}({\bf H})$ enlarges with increasing $N$, for a fixed desired level of multi-user interference energy, the
%symbol energy $E_k (k=1,2,\cdots,M$) can be increased with increasing $N$.
%
%Since, with increasing $N$ and fixed $E_k\,,\,k=1,2,\cdots,M$, an arbitrarily small multi-user interference energy can be guaranteed at each user,
%it can be argued that, with increasing $N$ the information symbol energy $E_k\,,\,k=1,2,\cdots,M$ can be increased while maintaining a fixed multi-user interference energy level at each user.
It is clear that, for a fixed $M$ and $N$, increasing $E_k,k=1,\ldots,M$ would enlarge ${\mathcal U}$ which could
then increase MUI energy level at each user (enlarging ${\mathcal U}$ might result in ${\mathcal U} \notin {\mathcal M}({\bf H})$). However, since an increase in $N$ (with fixed $M$ and $E_k$) results in a reduction of MUI (Theorem \ref{main_theorem1}),
it can be argued that for a fixed $M$, with increasing $N$ the information symbol energy
of each user (i.e., $E_k \,,\,k=1,\cdots,M$) can be increased while maintaining a fixed MUI energy level at each user.
Further, from (\ref{yk_model}), it is clear that for a fixed $P_T$ the effective SINR at the $k$-th user
(i.e., $E_k / ( {\mathbb E}_{{\bf u}}[\vert s_k \vert^2] \, + \, \sigma^2/P_T)$) will increase with increasing $N$, since $E_k$
can be increased while maintaining a constant MUI energy.
Finally, since $\sigma^2/P_T$ increases with decreasing $P_T$ and the MUI energy $\vert s_k \vert^2$ is independent of $P_T$, by appropriately {\em decreasing $P_T$ and increasing $E_k$ with
increasing $N$} (fixed $M$), a constant SINR level can be maintained at each user.
%Since the energy of the useful signal term (i.e., $\sqrt{P_T} \sqrt{E_k} u_k $) in the received signal $y_k$ (see (\ref{yk_model}))
%is $P_T E_k$, by decreasing $P_T$ with increasing $N$ (in such a way that $P_T E_k$ is constant) a constant SINR can be maintained at each user.

\begin{figure}[t]
\begin{center}
\epsfig{file=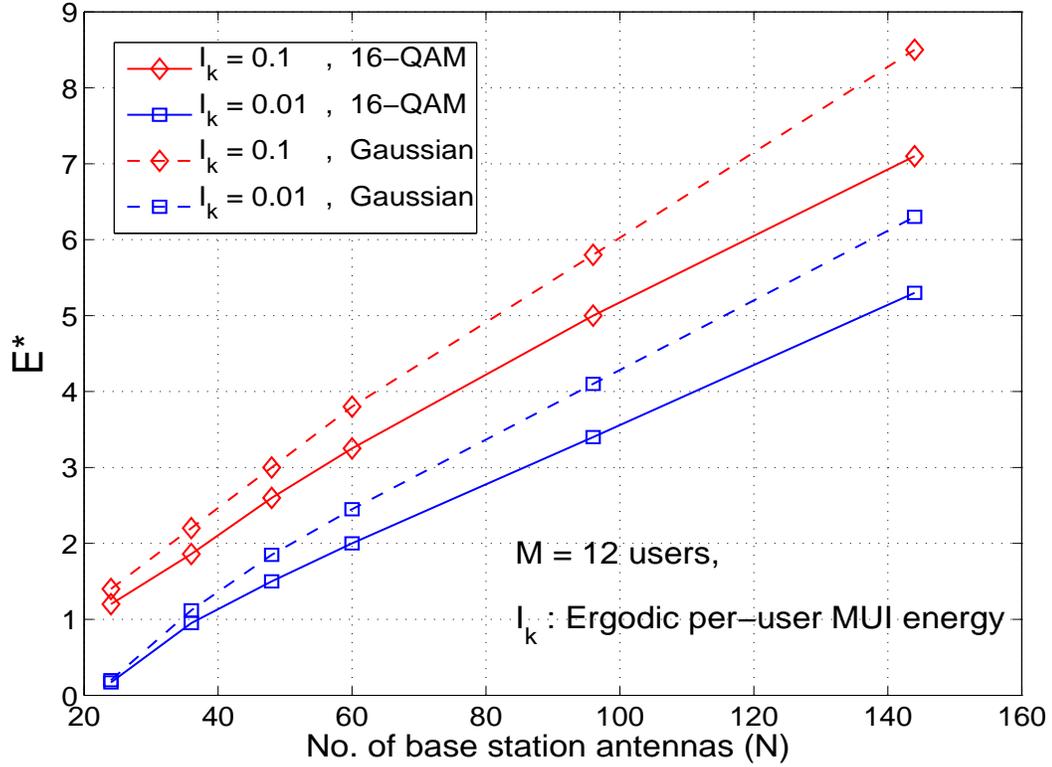, width=150mm,height=110mm}
\end{center}
\vspace{-6mm}
\caption{$E^\star$ vs. $N$ for a fixed desired ergodic MUI energy level $I_k$ (same for each user). Fixed $M=12$, fixed ${\mathcal U}_1=\cdots={\mathcal U}_M=\mbox{$16$-QAM},\mbox{Gaussian}$. IID ${\mathcal C}{\mathcal N}(0,1)$ Rayleigh fading.}
\label{fig_1}
\vspace{-5mm}
\end{figure}

This observation is based entirely on Theorem 1 (which holds for a broad class of fading channels satisfying
the conditions in (\ref{assumptions}), including i.i.d. fading channels).
%This then implies
%that, as long as the channel satisfies the conditions in (\ref{assumptions}), the achievable array gain will be greater than $1$ when $N$ is sufficiently large.
The above observation implies that as long as the channel satisfies the conditions in (\ref{assumptions}), 
the total transmit power can be reduced without affecting user information rates, by using a sufficiently large antenna array at the BS with constant
envelope transmission (i.e., {\em an achievable array gain greater than one}).
%In Section \ref{inc_sec}, through simulations, we show that indeed, for the i.i.d. Rayleigh fading channel with fixed $M$, $E_k$ can be increased linearly with increasing $N$,
%while still maintaining a fixed MUI energy level at each user.
We illustrate this through the following example using the proposed CE precoding scheme.
Let the fixed desired ergodic MUI energy level for the $k$-th user be denoted by
$I_k\,,\,k=1,2,\cdots,M$. For the sake of simplicity we consider ${\mathcal U}_1 = {\mathcal U}_2 = \cdots = {\mathcal U}_M$.
Consider
{
%\vspace{-3mm}
\begin{eqnarray}
\label{e_star}
E^\star \Define  {\mathop {\mbox{max}} \limits_{{p > 0 \,\,{\big |}\,\, E_k = p \,,\, {\mathbb E}_{{\bf H}}{\big [}{\mathbb E}_{_{u_1,\cdots,u_M}}{\big [}
{\vert} {\widehat s_k}  {\vert}^2 {\big ]}{\big ]} \, = \, I_k \,\,,\,\,k=1,\cdots,M}} }  p
\end{eqnarray}
}
which finds the highest possible equal energy of the information symbols under the constraint that the ergodic MUI
energy level is fixed at $I_k\,,\,k=1,2,\cdots,M$.
%The fact that we choose same information symbol energy for each user is not restrictive since all users have the same
%information alphabet and channel fading is i.i.d. Rayleigh.
In (\ref{e_star}),
${\widehat s_k}$ is given by (\ref{yk_model_hat}).
In Fig.~\ref{fig_1}, for the i.i.d. Rayleigh fading channel,
for a fixed $M = 12$ and a fixed ${\mathcal U}_1=\cdots={\mathcal U}_M=(\mbox{16-QAM and Gaussian})$,
we plot $E^\star$ as a function of increasing $N$, for two different fixed desired MUI energy
levels, $I_k= 0.1$ and $I_k=0.01$ (same $I_k$ for each user\footnote{\footnotesize{
Due to same channel gain distribution and information alphabet for each user,
it is observed that the ergodic MUI energy level at each user is also same if the users have equal information symbol energy.}}).
From Fig.~\ref{fig_1}, it can be observed that for a fixed $M$ and fixed ${\mathcal U}_1,\cdots,{\mathcal U}_M$,
$E^\star$ increases linearly with increasing $N$, while still maintaining a fixed MUI energy level at each user.
At low MUI energy levels, from (\ref{sinr_eq}) it follows that $\gamma_k \approx P_T E_k /\sigma^2$.
Since $E_k$ ($k=1,2,\cdots,M$) can be increased linearly with $N$ (while still maintaining a low MUI level),
it can be argued that a desired fixed SINR level can be maintained at each user by simply {\em reducing}
$P_T$ linearly with increasing $N$.
%This then suggests that, for a fixed $M$, an array power gain of $O(N)$ can be achieved under the stringent per-antenna CE constraint.
%In Section \ref{apg}, we confirm this observation on the achievability of an $O(N)$ array power gain.
This suggests the achievability of an $O(N)$ array power gain for the i.i.d. Rayleigh fading channel.
%For a given ${\bf H}$,
%,$E = (E_1,E_2,\cdots,E_M)^T$,
%and Gaussian information alphabets, it can be shown that 
%under the per-antenna CE constraint an information rate of
%$\log_2(\gamma_k({\bf H},E))$ is achievable for the $k$-the user ($\gamma_k({\bf H},E)$ is given by (\ref{sinr_eq})).
%\begin{eqnarray}
%\label{mi_expr}
%R_k({\bf H},E) & \Define & I(r_k ; {u_k}) \geq \log_2(\gamma_k({\bf H},E))
%\end{eqnarray}
%for each $k=1,2,\cdots,M$, where $\gamma_k({\bf H},E)$ is given by (\ref{sinr_eq}).
%With $ \log_2(\gamma_k({\bf H},E))$ as an achievable ergodic rate for the $k$-th user,
In the next section we derive an achievable sum-rate for the proposed CE precoding scheme,
using which (in Section \ref{sim_res}), for an i.i.d. Rayleigh fading channel, through simulations we show that indeed an $O(N)$ array power gain can be achieved.
{
%\vspace{-10mm}
}
\section {Achievable information sum rate}
In this section we study the ergodic information sum-rate achieved by the CE precoding scheme proposed in Section \ref{ce_prec_sec}.
For a given channel realization ${\bf H}$, Gaussian information alphabets\footnote{\footnotesize{We restrict the discussion to Gaussian information alphabets, due to the difficulty in analyzing the information rate achieved with discrete alphabets.
This is not a concern since, through Figs.~\ref{fig_0} and \ref{fig_1}, we have already observed that the two important results in Section \ref{remark1_sec} and \ref{remark2_sec} hold
true for Gaussian alphabets as well.}}$^,$\footnote{\footnotesize{We would also like to mention here that Gaussian information alphabets need not be optimal w.r.t. achieving the maximum sum-rate of a per-antenna CE constrained GBC.
As an example, in \cite{arxiv_su}, we have considered the capacity of a {\em single-user} MISO channel with per-antenna CE constraints at the transmitter.
Due to the scenario in \cite{arxiv_su} being simpler compared to the multi-user scenario discussed here, in \cite{arxiv_su} we were able to
show that the optimal capacity achieving complex alphabet is discrete-in-amplitude and uniform-in-phase (DAUIP) (i.e., non-Gaussian).
However, since it appears that the analytical tools and techniques in \cite{arxiv_su} cannot be used to derive the optimal alphabet for the multiuser scenario, we 
restrict ourselves to Gaussian alphabets here. 
%Nevertheless, it suffices to consider Gaussian alphabets as a starting point, since our main goal here is to address the question on achievability of array power gain
%under per-antenna CE transmission.
}} ${\mathcal U}_1,\cdots,{\mathcal U}_M$, information symbol
energies $E_1,\cdots,E_M$ and total transmit power to receiver noise ratio $P_T/\sigma^2$, the mutual information between $y_k$ and $u_k$ is given by
{
%\vspace{-8mm}
\begin{eqnarray}
\label{mut_inf_k}
I(y_k ; u_k)  =  h(u_k) - h(u_k \,  \vert  \, y_k)  = h(u_k) - h{\Big (} u_k - \frac{y_k}{\sqrt{P_T}\sqrt{E_k}} \,  {\Big \vert}  \, y_k {\Big )} 
 \geq  h(u_k) - h{\Big (}u_k - \frac{y_k}{\sqrt{P_T}\sqrt{E_k}} {\Big )}
\end{eqnarray}
%\vspace{-7mm}
}
where $h(z)$ denotes the differential entropy of a continuous valued r.v. $z$. The inequality in (\ref{mut_inf_k}) follows from the fact that conditioning of a r.v. reduces its entropy.
Further, using (\ref{yk_model_hat}) in (\ref{mut_inf_k}) we have
{
%\small
\begin{eqnarray}
\label{mut_inf_k1}
I(y_k ; u_k)   & \geq & h(u_k)  - h{\Big (} \frac{{\widehat s_k}}{\sqrt{E_k}} + \frac{w_k}{\sqrt{P_T} \sqrt{E_k}} {\Big )}  = \log_2(\pi e) - h{\Big (} \frac{{\widehat s_k}}{\sqrt{E_k}} + \frac{w_k}{\sqrt{P_T} \sqrt{E_k}} {\Big )} \nonumber \\
& \geq & \log_2(\pi e)  -  \log_2{\Bigg (} \pi e \, \mbox{var} {\Big [} \frac{{\widehat s_k}}{\sqrt{E_k}} + \frac{w_k}{\sqrt{P_T} \sqrt{E_k}} {\Big ]} {\Bigg )} \nonumber \\
& \geq & \log_2(\pi e) - \log_2{\Bigg (} \pi e \,  {\mathbb E} {\Big [} \,  {\Big \vert } \, \frac{{\widehat s_k}}{\sqrt{E_k}} + \frac{w_k}{\sqrt{P_T} \sqrt{E_k}} \, {\Big \vert}^2 \,  {\Big ]} {\Bigg )} \nonumber \\
& = & \log_2(\pi e) - \log_2{\Bigg (} \pi e \,   {\Big [} \frac { {\mathbb E}[\vert {\widehat s_k} \vert^2]}{E_k} + \frac {\sigma^2}{P_T E_k} {\Big ]} {\Bigg )} \, = \, \log_2{\Big (}\gamma_k({\bf H},E,\frac{P_T}{\sigma^2}){\Big )} \nonumber \\
& = & R_k{\Big (} {\bf H}, E, \frac{P_T}{\sigma^2} {\Big )}
\end{eqnarray}
}
%\normalsize
where $R_k{\Big (} {\bf H}, E, \frac{P_T}{\sigma^2} {\Big )}  \Define 
\log_2{\Big (}\gamma_k({\bf H},E,\frac{P_T}{\sigma^2}){\Big )}$
is an achievable information rate for the $k$-th user, with the proposed CE precoding scheme.
In (\ref{mut_inf_k1}), we have used the fact that the differential entropy of a complex Gaussian circular symmetric r.v. $z$ having variance $\sigma_z^2$ is
$\log_2(\pi e \sigma_z^2) $.
Further, for any complex scalar r.v. $z$, $\mbox{var}[z] \Define {\mathbb E} [ \vert z - {\mathbb E}[z] \vert^2 ]$.
The second inequality in (\ref{mut_inf_k1}) follows from the fact that, for a complex scalar r.v.,
among all possible probability distributions having the same variance, the complex circular symmetric Gaussian distribution is the entropy maximizer \cite{Cover}.
The third inequality follows from the fact that, for any complex scalar r.v. $z$, $\mbox{var}[z] \leq {\mathbb E}[ \vert z \vert^2]$.
%$\gamma_k({\bf H},E,\frac{P_T}{\sigma^2})$ is given by (\ref{sinr_eq}) and denotes the SINR for a given
%${\bf H}$, $E= (E_1,E_2,\cdots,E_M)^T$ and $P_T/\sigma^2$.
From (\ref{mut_inf_k1}) it follows that an achievable ergodic information sum-rate for the GBC under the per-antenna CE constraint, is given by
\begin{eqnarray}
\label{erg_sum_rate}
R^{\footnotesize \mbox{CE}}{\Big (}E,\frac{P_T}{\sigma^2} {\Big )} & \Define & \sum_{k=1}^M {\mathbb E}_{{\bf H}} {\Big [} \,  R_k{\Big (} {\bf H}, E, \frac{P_T}{\sigma^2} {\Big )} \, {\Big ]}.
\end{eqnarray}
Subsequently, we consider the scenario where all users have the same unit energy Gaussian information alphabet (i.e., ${\mathcal U}_1 = \cdots
= {\mathcal U}_M$) and the same information symbol energy (i.e.,
$E_1=E_2 = \cdots = E_M$).\footnote{ \footnotesize {We impose this constraint so as to reduce the number
of parameters involved, thereby simplifying the study of achievable rates in a multi-user GBC with per-antenna CE transmission.
Nevertheless, for the i.i.d. Rayleigh fading channel with each user having the same Gaussian information alphabet, it is expected that the optimal $E$ which maximizes the ergodic sum-rate in (\ref{erg_sum_rate}), has equal components.}}
Further optimization of $R^{\footnotesize \mbox{CE}}{\Big (}E,\frac{P_T}{\sigma^2} {\Big )}$ over $E$ subject to
$E_1=\cdots=E_M$, results in an achievable ergodic information sum-rate which is given by
\begin{eqnarray}
\label{sum_rate_final}
R^{\footnotesize \mbox{CE}}{\Big (} \frac{P_T}{\sigma^2} {\Big )} & \Define &   {\mathop  {\max} \limits_{E \, \vert \, E_1 = E_2 = \cdots = E_M > 0 } } R^{\footnotesize \mbox{CE}}{\Big (}E,\frac{P_T}{\sigma^2} {\Big )} 
\end{eqnarray}
Since it is difficult to analyze the sum-rate expression in (\ref{sum_rate_final}), we have studied it
through exhaustive numerical simulations for an i.i.d. ${\mathcal C}{\mathcal N}(0,1)$ Rayleigh fading channel.
In the following section, we present some important observations based on these numerical experiments.
{
\vspace{-4mm}
}
\section{Simulation results on the achievable ergodic information sum-rate $R^{\footnotesize \mbox{CE}}{\Big (} \frac{P_T}{\sigma^2} {\Big )}$}
\label{sim_res}
\begin{figure}[t]
\begin{center}
\epsfig{file=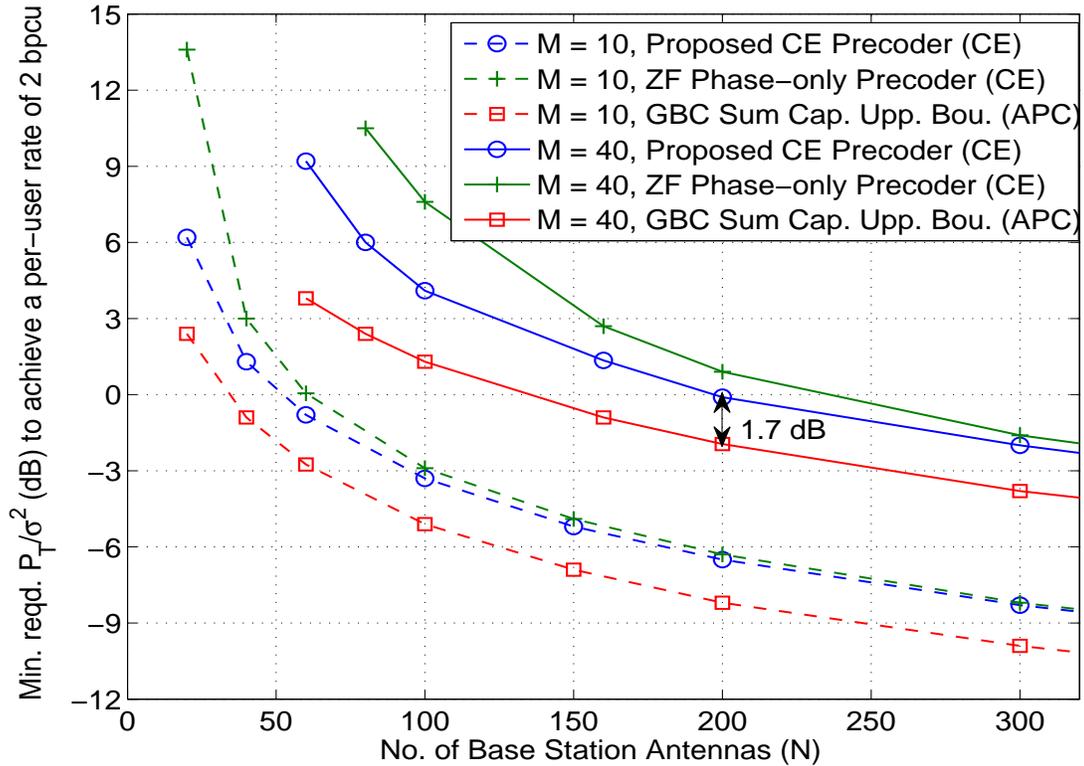, width=150mm,height=110mm}
\end{center}
{\vspace{-6mm}}
\caption{Required $P_T/\sigma^2$ vs. $N$, to achieve a fixed desired ergodic per-user rate $=2$ bpcu. Gaussian information alphabets ${\mathcal U}_1=\cdots={\mathcal U}_M$. IID ${\mathcal C}{\mathcal N}(0,1)$ Rayleigh fading.}
{\vspace{-4mm}}
\label{fig_2}
{\vspace{-2mm}}
\end{figure}

All reported results are for the i.i.d. ${\mathcal C}{\mathcal N}(0,1)$ Rayleigh fading channel.
In Fig.~\ref{fig_2}, for a fixed $M$
we plot the minimum $P_T/\sigma^2$ required by the proposed CE precoder, to achieve an ergodic per-user
information rate of $R^{\footnotesize \mbox{CE}}(P_T/\sigma^2)/M = 2$ bits-per-channel-use (bpcu) as a function of
increasing $N$ (Due to the same channel distribution for each user, we have observed that the ergodic information rate achieved by each user is $1/M$ of the ergodic sum-rate). The minimum required $P_T/\sigma^2$ is also tabulated in Table \ref{table_1}.
It is observed that, for a fixed $M$, at sufficiently large $N$, the required $P_T/\sigma^2$
reduces by roughly $3$ dB for every doubling in $N$
(i.e., the required $P_T/\sigma^2$ reduces linearly with increasing $N$).
This shows that, for a fixed $M$, an array power gain of $O(N)$ can indeed be achieved even under the stringent
per-antenna CE constraint.
For the sake of comparison, we have also plotted a lower bound on
the $P_T/\sigma^2$ required to achieve a per-user ergodic rate of $2$ bpcu under the APC constraint (we have used the cooperative upper bound on the GBC sum-capacity \cite{SV}).\footnote{\footnotesize{
The cooperative upper bound on the GBC sum capacity gives a lower bound on the $P_T/\sigma^2$ required
by a GBC sum-capacity achieving scheme to achieve a given desired ergodic information sum-rate.}}
We observe that, for large $N$ and a fixed per-user desired ergodic information rate of $2$ bpcu, compared to the APC only constrained GBC,
the extra total transmit power (power gap) required under the more stringent per-antenna CE constraint is {\em small} ($1.7$ dB).

In Fig.~\ref{fig_2}, we also consider another CE precoding scheme, where, for a given information symbol vector ${\bf u}$, the
precoder firstly computes the zero-forcing (ZF) vector ${\bf x} = {\bf H}^\dagger {\bf u}$, (${\bf H}^\dagger \Define {\bf H}^H {\Big (} {\bf H} {\bf H}^H  {\Big )}^{-1}$ is the pseudo-inverse of ${\bf H}$).
Prior to transmission, each component of ${\bf x}$ is normalized to have a modulus equal to $\sqrt{P_T/N}$,
i.e., the signal transmitted from the $i$-th BS antenna is $\sqrt{P_T/N} \, x_i/\vert x_i \vert$. At each user, the
received signal is scaled by a fixed constant.\footnote{\footnotesize{This constant is chosen in such a way that the ergodic per-user information rate is maximized. It is therefore fixed for all channel realizations and depends only upon the statistics of the channel,
$P_T/\sigma^2$, $N$ and $M$.}}
We shall hence-forth refer to this precoder as the ZF phase-only precoder.
In Fig.~\ref{fig_2}, we observe that the $P_T/\sigma^2$ required by the proposed CE precoder is always less
than that required by the ZF phase-only precoder. 
In fact, for moderate values of $N/M$, the proposed CE precoder requires significantly less $P_T/\sigma^2$ as compared to the
 ZF phase-only precoder (e.g. with $N=100,M=40$, the required $P_T/\sigma^2$ with the proposed CE precoder is roughly $3$ dB less
than that required with the
ZF phase-only precoder). 
However, at very large values of $N/M$, the ZF phase-only precoder has similar performance as the
proposed CE precoder.\footnote{\footnotesize{Note that the ZF phase-only precoder does not necessarily have a lower complexity
than the proposed CE precoder.
This is because, the ZF phase-only precoder needs to compute the pseudo-inverse of the channel gain matrix (a $M \times N$ matrix) and also the matrix vector product of the pseudo-inverse times the information symbol vector ${\bf u}$. Computing the pseudo-inverse
has a
complexity of $O(M^2 N)$ and that for the matrix vector product is $O(MN)$, resulting in a total complexity of $O(M^2 N)$ for the ZF phase-only precoder.
In contrast, the proposed CE precoder does not need to compute the pseudo-inverse, and has a complexity of $O(M N)$
(see Section \ref{ce_prec_sec}) as compared to the $O(M^2 N)$ complexity of the ZF phase-only precoder.}}

\begin{figure}[t]
\begin{center}
\epsfig{file=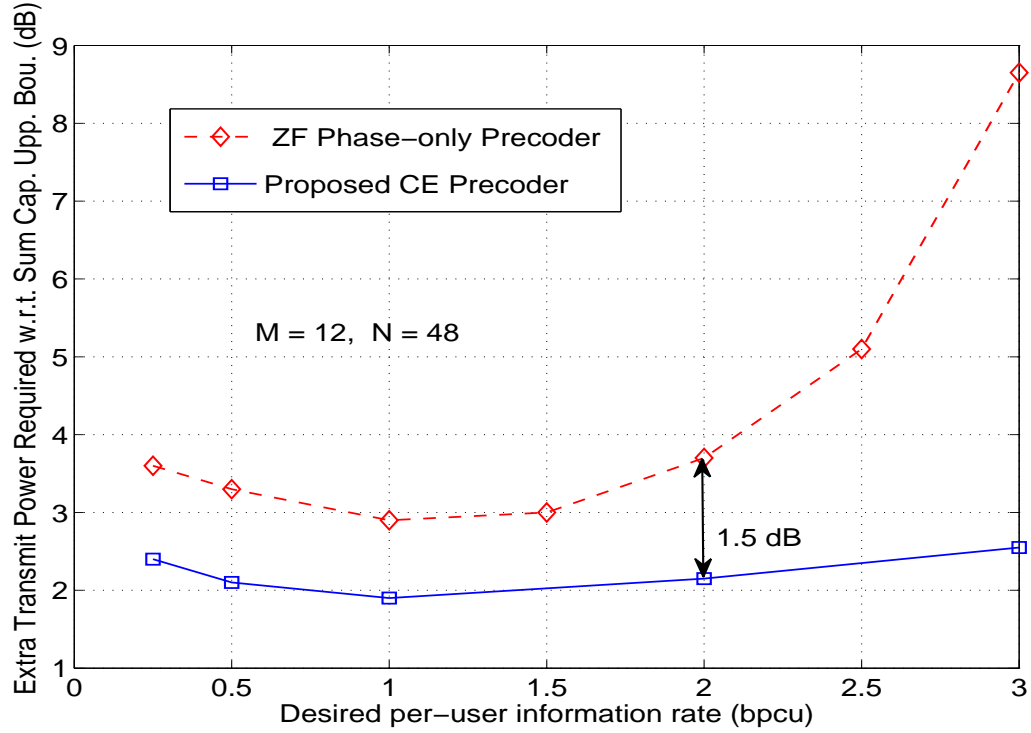, width=150mm,height=100mm}
\end{center}
{\vspace{-6mm}}
%\caption{Fixed $N = 48, M = 12$. Compared to a GBC sum capacity achieving scheme (under APC), an upper bound on the extra $P_T/\sigma^2$ (dB) required by the proposed CE precoder and the ZF phase-only precoder is plotted as a function of the desired per-user ergodic information rate. Gaussian information alphabets ${\mathcal U}_1=\cdots={\mathcal U}_M$. IID ${\mathcal C}{\mathcal N}(0,1)$ Rayleigh fading.}
\caption{The extra $P_T/\sigma^2$ (in dB) required (vertical axis) by the proposed CE precoder and by the ZF phase-only precoder, respectively, to achieve the same
ergodic per-user information rate as predicted by the GBC sum-capacity cooperative upper bound (horizontal axis). Here the number of base station antennas is $N = 48$ and the number of users is $  M = 12$.  All users use Gaussian information alphabets ${\mathcal U}_1=\cdots={\mathcal U}_M=\mbox{Gaussian}$ and all channels are  i.i.d. ${\mathcal C}{\mathcal N}(0,1)$ Rayleigh fading.}
{\vspace{-4mm}}
\label{fig_5}
{\vspace{-2mm}}
\end{figure}

\begin{figure}[t]
\begin{center}
\epsfig{file=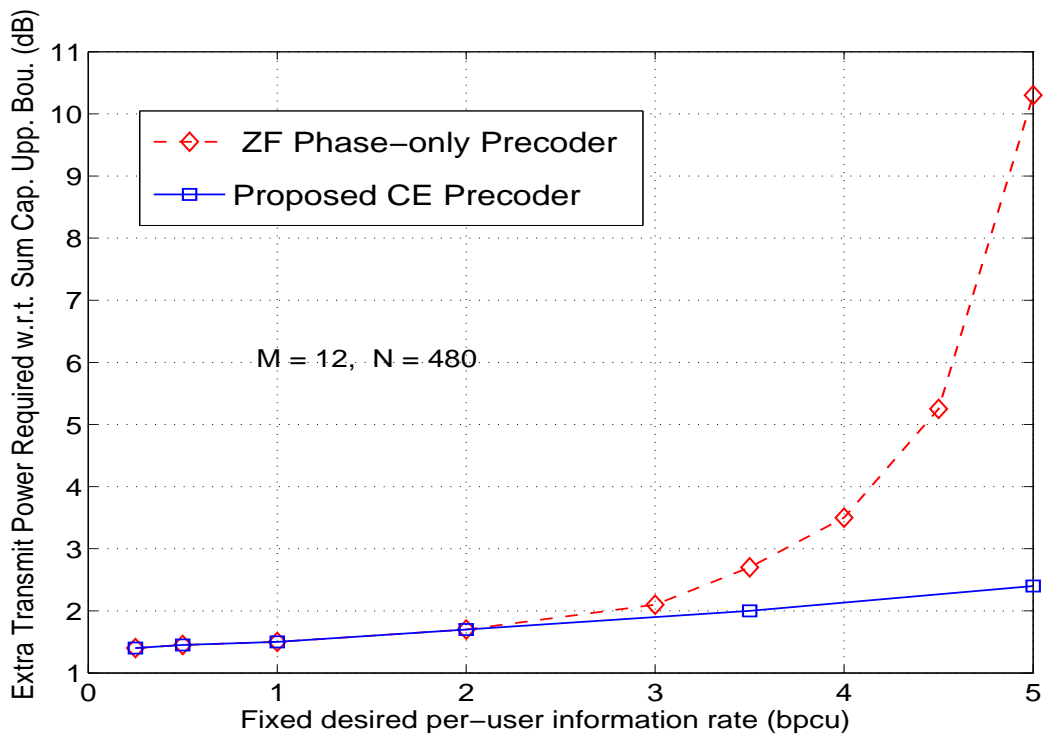, width=150mm,height=100mm}
\end{center}
{\vspace{-6mm}}
\caption{Same as Fig.~\ref{fig_5}, but for $N=480$ base station antennas.}
{\vspace{-4mm}}
\label{fig_6}
{\vspace{-2mm}}
\end{figure}

To gain a better understanding of the power-efficiency of the considered CE precoders, in Fig.~\ref{fig_5},
for a fixed $N=48, M=12$ we plot an upper bound on the extra $P_T/\sigma^2$ required by the considered CE precoding schemes when compared to a GBC sum-capacity achieving
scheme under APC,\footnote{\footnotesize{
Since we use the cooperative upper bound to predict the $P_T/\sigma^2$ required by a GBC sum-capacity achieving
scheme, the reported values of the extra $P_T/\sigma^2$ required by the considered CE precoders
are infact an upper bound on the minimum extra $P_T/\sigma^2$ required.}}
as a function of the desired per-user ergodic information rate (note that in Fig.~\ref{fig_2}, the
desired per-user rate was fixed to $2$ bpcu). It is observed that, for a desired ergodic per-user information rate below $2$ bpcu, the
ZF phase-only precoder requires roughly $1-1.5$ dB more transmit power as compared to the proposed CE precoder.
For rates higher than $2$ bpcu, this gap increases very rapidly (at $3$ bpcu, this power gap is roughly $6$ dB).
In Fig.~\ref{fig_6}, we plot the results of a similar experiment but with $N=480, M=12$ (a very large ratio of $N/M$).
It is observed that, the ZF phase-only precoder has similar performance as the proposed CE precoder for per-user ergodic information
rates below $3$ bpcu. For rates higher than $3$ bpcu, the performance of the ZF phase-only precoder deteriorates rapidly, just
as it did in Fig.~\ref{fig_5}.
In Figs.~\ref{fig_5} and \ref{fig_6}, we also note that the extra total transmit power required
by the proposed CE precoder (Section \ref{ce_prec_sec}) increases slowly w.r.t. increasing rate, and is less than $2.5$ dB for a wide range of desired per-user information rates.
From exhaustive experiments, we have concluded that, for moderate values of $N/M$, the proposed CE precoder
is significantly more power efficient than the ZF phase-only precoder, whereas for very large $N/M$ both precoders have similar
performance when the desired per-user ergodic information rate is below a certain threshold (beyond this threshold, the performance of the ZF phase-only
precoder deteriorates).

\begin{figure}[t]
\begin{center}
\epsfig{file=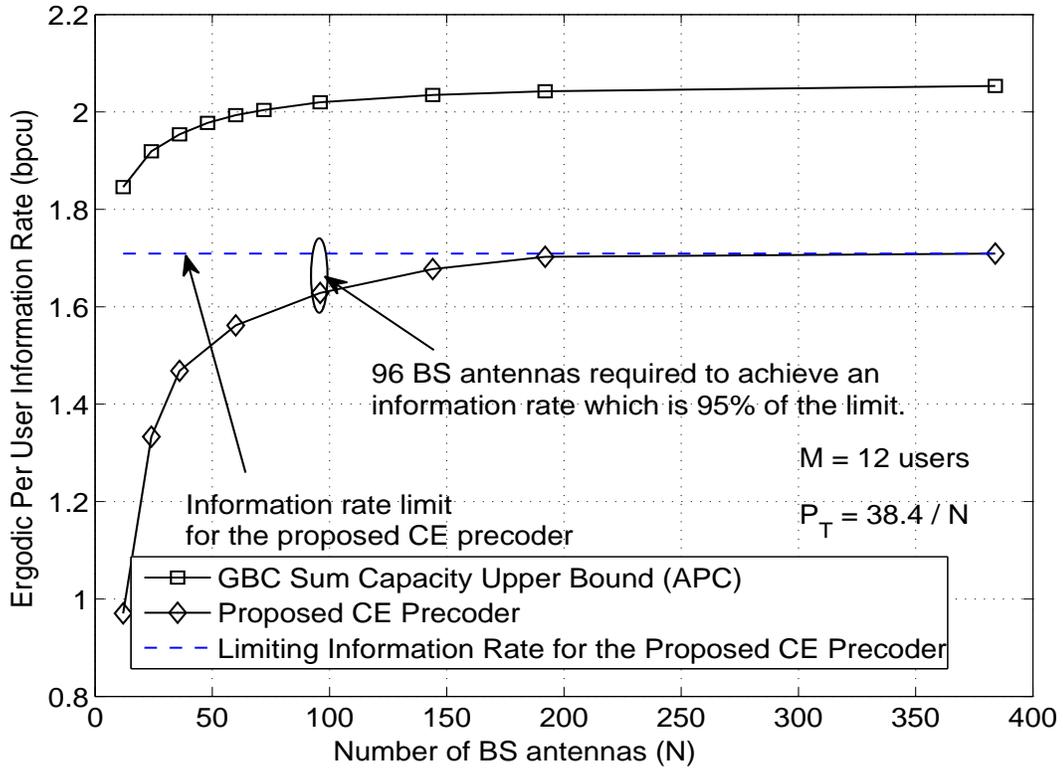, width=150mm,height=110mm}
\end{center}
{\vspace{-6mm}}
\caption{Ergodic per-user information rate for a fixed $M=12$, with the total transmit power scaled down
linearly with increasing $N$. Gaussian information alphabets ${\mathcal U}_1=\cdots={\mathcal U}_M$. IID ${\mathcal C}{\mathcal N}(0,1)$ Rayleigh fading.}
{\vspace{-4mm}}
\label{fig_3}
{\vspace{-2mm}}
\end{figure}

\begin{figure}[t]
\begin{center}
\epsfig{file=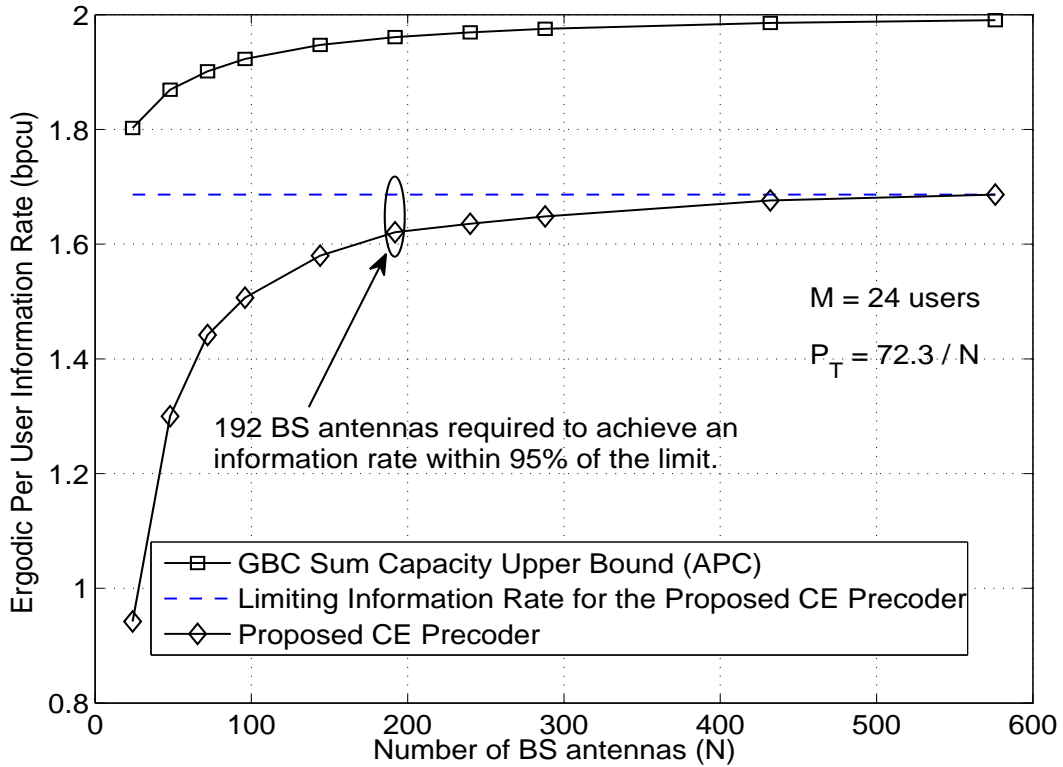, width=150mm,height=110mm}
\end{center}
{\vspace{-6mm}}
%\caption{Ergodic per-user information rate for a fixed $M=24$, with the total transmit power scaled down
%linearly with increasing $N$. Gaussian information alphabets ${\mathcal U}_1=\cdots={\mathcal U}_M$. IID ${\mathcal C}{\mathcal N}(0,1)$ Rayleigh fading.}
\caption{Same as Fig.~\ref{fig_3}, but with a fixed $M=24$ and $P_T=72.3/N$.}
\label{fig_4}
\end{figure}

In Fig.~\ref{fig_2}, for the proposed CE precoder, we had observed that for a fixed $M$ and fixed desired per-user information rate,
with ``sufficiently large'' $N$, the total transmit power can be reduced linearly with increasing $N$.
We next try to understand as to how ``large'' must $N$ be, so that $P_T/\sigma^2$ can be reduced
by roughly $3$ dB with every doubling in $N$ (fixed $M$), while maintaining a fixed achievable per-user ergodic information rate.
In Fig.~\ref{fig_3}, for a fixed $M=12$ users, we plot the achievable per-user ergodic information rate under per-antenna
CE transmission
(i.e., $R^{\footnotesize \mbox{CE}}{\Big (} \frac{P_T}{\sigma^2} {\Big )}/M$) as a function
of increasing $N$ and $P_T = P_0/N$ (i.e., we linearly decrease $P_T$ with increasing $N$, $P_0 = 38.4$).
It is observed that, the per-user ergodic information rate increases and approaches a limiting information rate
as $N \rightarrow \infty$ (shown by the dashed curve in the figure). $P_0 = 38.4$ corresponds to a limiting
per-user information rate of roughly $1.7$ bpcu.
This then suggests that, in the limit as $N \rightarrow \infty$, the per-user information rate remains
fixed as long as $P_T$ is scaled down linearly with increasing $N$ (this re-confirms our conclusion
on the achievability of an $O(N)$ array power gain under the per-antenna CE constraint, for a i.i.d. Rayleigh fading channel).
A similar behaviour is observed under APC (see the GBC sum capacity upper bound curve in the figure).
In Fig.~\ref{fig_4}, similar results have been illustrated for $M = 24$ users and $P_T = P_1/N$ ($P_1 = 72.3$, corresponding to a
limiting
per-user information rate of roughly $1.7$ bpcu).
With regards to the question on how ``large'' must $N$ be, it is now clear that $N$ must at least be so large
that the achievable per-user ergodic information rate is sufficiently close to its limiting information rate (i.e., in the flat region of the curve).
In general, for a desired closeness\footnote{\footnotesize{Closeness could be expressed 
in terms of the achievable per-user ergodic information rate being greater than a specified percentage of the
limiting information rate.}} to the limiting information rate, the minimum number of 
BS antennas required depends on $M$. Our numerical experiments suggest that, to achieve a fixed desired ratio
of the per-user ergodic information rate to the limiting information rate, a channel with a large $M$
requires a large $N$ also.
%In fact our simulations suggest that, for $N > M \gg1 $, there exists a relation between
%the fixed desired ratio 
%of the achievable per-user ergodic information rate to the limiting information rate, and the ratio of the 
%minimum required number of base station antennas to $M$.
As an example, for a fixed ratio of $0.95$ between the achievable per-user ergodic information rate and
the limiting information rate, a channel with $M=12$ users requires a BS with at least $N=96$ antennas,
whereas a channel with $M=24$ users requires a BS with at least $N=192$ antennas (i.e., to achieve
an ergodic per-user information rate within $95$ percent of the limiting information rate requires
a BS with roughly $8$ times more number of antennas than the number of users).
{
\vspace{-7mm}
}
\section{Conclusion}
In this paper, we have considered per-antenna constant envelope (CE) transmission in the downlink of
multi-user MIMO systems (GBC) employing a large number of BS antennas.
Under certain mild conditions on the channel, even with a stringent per-antenna CE constraint, array power gain can still
be achieved.
We have also proposed a low-complexity CE precoding scheme.
For the proposed CE precoding scheme, through exhaustive simulations for the i.i.d. Rayleigh fading channel, it is shown that, compared to an APC only constrained GBC,
the extra total transmit power required by the proposed CE precoder to achieve a given per-user ergodic information rate is small
(usually less than $2$ dB for scenarios of interest).
%Due to the extra total transmit power required being small, it follows that, in terms of the overall power-efficiency of the base station,
%it would be better to have power-efficient non-linear amplifiers with CE inputs instead of
%having highly linear but power-inefficient amplifiers with high PAPR inputs.
Typically, a non-linear power-efficient amplifier is about $4-6$ times more power-efficient than a highly linear amplifier \cite{cripps}.
%With a sufficiently large $N$, from the above information-theoretic results we know that,
%to achieve a given information sum-rate, CE inputs would require
%roughly about $1.7$ dB extra total radiated power as compared to average total power only constrained inputs.
Combining this fact with the fact that per-antenna CE signals require an extra $2$ dB transmit power, we arrive at the conclusion that,
for a given desired achievable information sum-rate, with sufficiently large $N$, a base station having power-efficient amplifiers with CE inputs would require
$10 \log_{10}(4) - 2.0 = 4.0$ dB lesser total transmit power compared to a base station having highly linear
power-inefficient amplifiers with high PAPR inputs.
%\bibliographystyle{IEEEbib}
%\bibliography{strings,refs}
{
\vspace{-5mm}
}

{
\vspace{-6mm}
}
\appendices
%\section{Proof of Theorem \ref{main_theorem1} }
%\label{dim_sec}
\section{Convergence (in distribution) of the sequence $ {\Big \{} {\bf z}_N {\Big \}}$ }
\label{app_conv_zN}
The convergence in distribution of the sequence of random variables $ {\Big \{} {\bf z}_N {\Big \}}$ (as $N \rightarrow \infty$ with fixed $M$) is stated
and proved in Theorem \ref{thm_conv_zN}.
Its proof relies on three known results which have been stated below.

\begin{myresult}\label{multivariate_clt}
(Multivariate Central Limit Theorem (CLT))
Let $F_n$ denote the joint cumulative distribution function (c.d.f.) of the $k$-dimensional real random variable $(X_n^{(1)},\cdots,X_n^{(k)})$, $n=1,2,\ldots$ and
for each real vector ${\boldsymbol \Lambda} = (\lambda_1, \lambda_2, \cdots, \lambda_k)^T$, let $F_{{\boldsymbol \Lambda} \, n}$
be the c.d.f. of the random variable $\lambda_1 X_n^{(1)} + \lambda_2 X_n^{(2)} + \cdots + \lambda_k X_n^{(k)}$.
A necessary and sufficient condition for $F_n$ to converge to a limiting distribution (as $n \rightarrow \infty$) is that
$F_{{\boldsymbol \Lambda} \, n}$ converges to a limit for {\em each} vector ${\boldsymbol \Lambda}$.
\end{myresult}

{\it Proof} --
For details please refer to \cite{Varadarajan}
. $\hfill\blacksquare$

This result basically states that, if $F$ is the joint c.d.f. of a $k$-dimensional real
random variable $(X^{(1)}, X^{(2)}, \cdots, X^{(k)})$, and if $F_{{\boldsymbol \Lambda} \, n} \rightarrow F_{{\boldsymbol \Lambda}}$ 
for\footnote{\footnotesize{
$F_{{\boldsymbol \Lambda}}
$ is the c.d.f. of $\lambda_1 X^{(1)} + \cdots + \lambda_k X^{(k)}$.}} each vector ${\boldsymbol \Lambda}$, then
$F_n \rightarrow F$ as $n \rightarrow \infty$.

\begin{myresult}\label{lyapunov_th}
(Lyapunov-CLT) Let $\{X_n\}$, $n=1,2,\ldots$ be a sequence of independent real-valued scalar random variables.
Let ${\mathbb E}[X_n] = \mu_n$, ${\mathbb E}[ (X_n - \mu_n)^2] = \sigma_n^2 $,
and for some fixed $\xi > 0$,
${\mathbb E}[ \vert X_n - \mu_n \vert^{2 + \xi}] = \beta_n$ exists for all $n$.
Furthermore let
{
\small
\begin{equation}
\label{lyapunov_def}
B_n \Define {\Big (} \sum_{i=1}^n \beta_i {\Big )}^{\frac{1}{2 + \xi}}\,\,,\,\, C_n \Define {\Big (} \sum_{i=1}^n \sigma_i^2 {\Big )}^{\frac{1}{2 }}.
\end{equation}
}
\normalsize
Then if
{
\vspace{-8mm}
}
{
\small
\begin{equation}
\label{lyapunov_cnd1}
\lim_{n \rightarrow \infty} \frac{B_n}{C_n} = 0,
\end{equation}
}
\normalsize
the c.d.f. of
%\begin{equation}
%\label{lyapunov_stmt1}
$Y_n = \frac{\sum_{i=1}^n (X_i - \mu_i) }{C_n}$
%\end{equation}
converges (in the limit as $n \rightarrow \infty$) to the c.d.f. of a real Gaussian random variable with mean zero and unit variance.
\end{myresult}

{\it Proof} --
For details please refer to \cite{Patrick}
. $\hfill\blacksquare$

\begin{myresult}\label{slutsky} (Slutsky's Theorem)
Let $\{ X_n \}$ and $\{ Y_n \}$ be a sequence of scalar random variables.
If $\{ X_n \}$ converges in distribution (as $n \rightarrow \infty$) to some random variable $X$, and $\{ Y_n \}$ converges
in probability to some constant $c$, then the product sequence
$\{ X_n Y_n \}$ converges in distribution to the random variable $c X$.
\end{myresult}

{\it Proof} --
For details please refer to \cite{Basu}
. $\hfill\blacksquare$

\begin{mytheorem}\label{thm_conv_zN}
For any channel sequence $\{ {\bf H}_N \}$ satisfying the conditions in (\ref{assumptions}),
the associated sequence of random vectors $\{ {\bf z}_N \} $ (defined in (\ref{zN_def}))
converges (as $N \rightarrow \infty$ with fixed $M$) in distribution to a multivariate $2M$-dimensional real Gaussian random vector
$X = (X_1^I,X_1^Q,\cdots,X_M^I,X_M^Q)^T$
with independent zero-mean
components and $\mbox{var}(X_k^I) = \mbox{var}(X_k^Q) = c_k/2 \,,\,k=1,2,\ldots,M$ (note that
$c_k \,,\,k=1,2,\ldots,M$ is defined in (\ref{assumptions})).
\end{mytheorem}

{\it Proof} --
%We basically use Result \ref{multivariate_clt} and
%Theorem \ref{linear_z} to prove Theorem \ref{thm_conv_zN}.
Consider a multivariate $2M$-dimensional real random variable $(X_1^I, X_1^Q, \cdots, X_{M}^I,X_M^Q)$, whose
components are i.i.d. real Gaussian with mean zero and $\mbox{var}(X_k^I) = \mbox{var}(X_k^Q) = c_k/2 \,,\,k=1,2,\ldots,M$.
Then, for any vector ${\boldsymbol \Lambda} = (\lambda_1^I, \lambda_1^Q, \cdots, \lambda_M^I, \lambda_M^Q)^T \in {\mathbb R}^{2M}$, the scalar random variable
$(\lambda_1^I X_1^I + \lambda_1^Q X_1^Q + \cdots + \lambda_M^I X_M^I + \lambda_M^Q X_M^Q)$ is real Gaussian with mean zero and variance
$\sum_{k=1}^M c_k {\big (} (\lambda_k^I)^{^2} + (\lambda_k^Q)^{^2} {\big )}/2$.

If we can show that for any arbitrary vector ${\boldsymbol \Lambda}  \in {\mathbb R}^{2M}$,
the limiting distribution of ${\bf z}_N^T {\boldsymbol \Lambda}$ is also
real Gaussian with mean zero and the same variance $\sum_{k=1}^M c_k {\big (} (\lambda_k^I)^{^2} + (\lambda_k^Q)^{^2} {\big )}/2$,
then using Result \ref{multivariate_clt} it will follow that the
c.d.f. of ${\bf z}_N$ converges to the c.d.f. of $(X_1^I, X_1^Q, \cdots, X_{M}^I,X_M^Q)$ as
$N \rightarrow \infty$. This would then complete the proof.

In the following we show that under the assumptions stated in (\ref{assumptions}), for any arbitrary vector ${\boldsymbol \Lambda}  \in {\mathbb R}^{2M}$,
{\em indeed} the limiting distribution (i.e., as $N \rightarrow \infty$ with fixed $M$) of ${\bf z}_N^T {\boldsymbol \Lambda}$ is real Gaussian with mean zero and variance $\sum_{k=1}^M c_k {\big (} (\lambda_k^I)^{^2} + (\lambda_k^Q)^{^2} {\big )}/2$, thereby completing the proof.

For a given $2M$-dimensional real vector ${\boldsymbol \Lambda} = (\lambda_1^I, \lambda_1^Q, \cdots, \lambda_M^I, \lambda_M^Q)^T$,
let
{
\small
\begin{eqnarray}
\label{def_zetaN}
\zeta_N \Define {\bf z}_N^T {\boldsymbol \Lambda} = \sum_{k=1}^M (\lambda_k^I {z_k^I}^{(N)} + \lambda_k^Q {z_k^Q}^{(N)}).
\end{eqnarray}
}
\normalsize
%Under the assumptions stated in (\ref{assumptions}), as $N \rightarrow \infty$ the c.d.f. of $\zeta_N$
%converges to the c.d.f. of a zero mean real Gaussian random variable with variance
%$\sum_{k=1}^M c_k {\big (} (\lambda_k^I)^{^2} + (\lambda_k^Q)^{^2} {\big )}/2$.
From the above definition and (\ref{zN_def}), it follows that r.v. $\zeta_N$ can be expressed as\footnote{\footnotesize{Note
that the randomness in ${\bf z}_N$ is only due to the random variables $\theta_i\,,\,i=1,2,\ldots,N$.}}
{
\small
\begin{eqnarray}
\label{zeta_N_def}
\zeta_N & = & \sum_{i=1}^N (a_i \cos(\theta_i) + b_i \sin(\theta_i)) 
 =  \sum_{i=1}^N \sqrt{a_i^2 + b_i^2} \cos(\theta_i - \tan^{-1}\frac{b_i}{a_i}) \nonumber \\
a_i & \Define & \frac{\sum_{k=1}^M (\lambda_k^I {h_{k,i}^{I^{(N)}}} + \lambda_k^Q {h_{k,i}^{Q^{(N)}}})} {\sqrt{N}} \,\,\,,\,\,\,
b_i  \Define  \frac{\sum_{k=1}^M (\lambda_k^Q {h_{k,i}^{I^{(N)}}} - \lambda_k^I {h_{k,i}^{Q^{(N)}}})} {\sqrt{N}} 
%h_{k,i}^{I^{(N)}} & \Define & \mbox{Re}(h_{k,i}^{(N)})\,\,,\,\, h_{k,i}^{Q^{(N)}} \, \Define \, \mbox{Im}(h_{k,i}^{(N)}).
\end{eqnarray}
}
\normalsize
where $h_{k,i}^{I^{(N)}}  \Define  \mbox{Re}(h_{k,i}^{(N)})\,\,,\,\, h_{k,i}^{Q^{(N)}} \, \Define \, \mbox{Im}(h_{k,i}^{(N)})$. We further define
{
\small
\begin{eqnarray}
\eta_i & \Define & \sqrt{a_i^2 + b_i^2} \cos(\theta_i - \tan^{-1}\frac{b_i}{a_i})
\end{eqnarray}
}
\normalsize
Since, the phase angles $\theta_i,i=1,2,\ldots,N$ are independent of each other, $\eta_i, i=1,2,\cdots,N$
are also independent.
Therefore, $\zeta_N$ is nothing but the sum of $N$ independent random variables.
We can therefore apply the Lyapunov-CLT (Result \ref{lyapunov_th}) to study the convergence of the c.d.f.
of $\zeta_N$ as $N \rightarrow \infty$.

We firstly see that $\mu_i \Define {\mathbb E}[\eta_i] = 0 $ and $\sigma_i^2 \Define {\mathbb E}[\eta_i^2] = (a_i^2 + b_i^2)/2 $ since $\theta_i$ is uniformly distributed in $[-\pi, \pi)$.
We next show that the conditions
of the Lyapunov-CLT ((\ref{lyapunov_cnd1}) in Result \ref{lyapunov_th}) are satisfied with $\xi = 2$.
We see that
{
\vspace{-3mm}
}
{
\small
\begin{eqnarray}
\beta_i \Define {\mathbb E}[\eta_i^4] & = & (a_i^2 + b_i^2)^2 {\mathbb E}[\cos^4(\theta_i - \tan^{-1}\frac{b_i}{a_i})] 
\,  =  \, \frac{3}{8} (a_i^2 + b_i^2)^2
\end{eqnarray}
}
\normalsize
exists for all $i$.
In order that the condition in (\ref{lyapunov_cnd1}) is satisfied, we must show that
{
\small
\begin{equation}
\label{th3_eq3}
\lim_{N \rightarrow \infty}  \frac{B_N}{C_N} = 0
\end{equation}
}
\normalsize
where
{
\vspace{-7mm}
}
{
\small
\begin{eqnarray}
\label{BC_N_def}
B_N & \Define & {\Big (} \sum_{i=1}^N \beta_i {\Big )}^{\frac{1}{4}} =  {\Big (} \frac{3}{8} \sum_{i=1}^N (a_i^2 + b_i^2)^2  {\Big )}^{\frac{1}{4}} \,\,,\,\,
C_N \, \Define \, {\Big (} \sum_{i=1}^N \sigma_i^2 {\Big )}^{\frac{1}{2}} =  {\Big (} \sum_{i=1}^N (a_i^2 + b_i^2)/2  {\Big )}^{\frac{1}{2}}
\end{eqnarray}
}
\normalsize
As a note, from (\ref{zeta_N_def}) it follows that both $B_N$ and $C_N$
are strictly positive for all $N \geq M$.
Since $M$ is fixed, proving (\ref{th3_eq3}) is therefore equivalent to proving that
{
\small
\begin{equation}
\label{th3_eq3_equiv}
\lim_{N \rightarrow \infty}  \frac{B_N^4}{C_N^4} = 0
\end{equation}
}
\normalsize
%We next prove (\ref{th3_eq3_equiv}).
Using (\ref{assumptions}) we firstly show that
{
\small
\begin{equation}
\label{stmt1_th3}
\lim_{N \rightarrow \infty} C_N^2 = \frac{1}{2} \sum_{k=1}^M c_k {\big (} (\lambda_k^I)^{2} + (\lambda_k^Q)^{2} {\big )}
\end{equation}
}
\normalsize
i.e., $C_N^2$ converges to a constant as $N \rightarrow \infty$.
We then show that, again under (\ref{assumptions}),
{
\small
\begin{equation}
\label{stmt2_th3}
\lim_{N \rightarrow \infty} B_N^4 = 0
\end{equation}
}
\normalsize
Equation (\ref{th3_eq3_equiv}) would then follow from (\ref{stmt1_th3}) and (\ref{stmt2_th3}).
We next show (\ref{stmt1_th3}).
Using (\ref{BC_N_def}) we have $2C_N^2 = \sum_{i=1}^N (a_i^2 + b_i^2)$. Expanding the expressions for $a_i$ and $b_i$ in $\sum_{i=1}^N (a_i^2 + b_i^2)$ using (\ref{zeta_N_def}), we have
{
\small
\begin{eqnarray}
\label{step1_th3}
2C_N^2 
& = & \sum_{k=1}^M (({\lambda_k^I})^2 + ({\lambda_k^Q})^2) \frac { \Vert {\bf h}_k^{(N)} \Vert^2  } {N} \,\,  + \,\,
2 \sum_{k=1}^M \sum_{l = k+1}^M {\Big \{} 
 (\lambda_k^I \lambda_l^I + \lambda_k^Q \lambda_l^Q) \frac{\sum_{i=1}^N (h_{k,i}^{I^{(N)}} h_{l,i}^{I^{(N)}} + h_{k,i}^{Q^{(N)}} h_{l,i}^{Q^{(N)}})}{N}  \nonumber  \\
& & \hspace{62mm}  + (\lambda_k^I \lambda_l^Q - \lambda_k^Q \lambda_l^I) \frac{\sum_{i=1}^N (h_{k,i}^{I^{(N)}} h_{l,i}^{Q^{(N)}} - h_{k,i}^{Q^{(N)}} h_{l,i}^{I^{(N)}})}
{N}  {\Big \}}.
\end{eqnarray}
}
\normalsize
From $\mbox{As.1}$ and $\mbox{As.3}$ in (\ref{assumptions}) it follows that
{
\small
\begin{eqnarray}
\label{step2_th3}
\lim_{N \rightarrow \infty} \frac{\sum_{i=1}^N (h_{k,i}^{I^{(N)}} h_{l,i}^{I^{(N)}} + h_{k,i}^{Q^{(N)}} h_{l,i}^{Q^{(N)}})}{N}   =  0\,\,,\,\,
\lim_{N \rightarrow \infty} \frac{\sum_{i=1}^N (h_{k,i}^{I^{(N)}} h_{l,i}^{Q^{(N)}} - h_{k,i}^{Q^{(N)}} h_{l,i}^{I^{(N)}})} 
{N} =  0 \,\,,\,\,
\lim_{N \rightarrow \infty}  \frac { \Vert {\bf h}_k^{(N)} \Vert^2  } {N}  =  c_k .
\end{eqnarray}
}
\normalsize
Using (\ref{step2_th3}) in (\ref{step1_th3}) and taking the limit as $N \rightarrow \infty$ we get
(\ref{stmt1_th3}) (note that $M$ is fixed).
We now show (\ref{stmt2_th3}).
Before proceeding further, we define the complex numbers $\lambda_k \Define (\lambda_k^I + j \lambda_k^Q), k=1,2,\ldots,M$.
%Also, $*$ is used to denote complex conjugation, i.e., $\lambda_k^* \Define \lambda_k^I - j \lambda_k^Q$.
Expanding the expressions for $a_i$ and $b_i$ inside the summation in $B_N^4$ (see (\ref{BC_N_def})) we have
{
\small
\begin{eqnarray}
\label{step3_th3}
\frac{8}{3} B_N^4 & = & \sum_{i=1}^N (a_i^2 + b_i^2)^2 \nonumber \\
& = & \sum_{i=1}^N {\Bigg \{}  \sum_{k=1}^M \frac {\vert \lambda_k \vert^2 \vert h_{k,i}^{(N)} \vert^2 } {N} 
\,\,  + \,\, 2 \sum_{k=1}^M\sum_{l = k+1}^M \frac { {\big (} \mbox{Re}(\lambda_k^* \lambda_l) \mbox{Re}( h_{k,i}^{{(N)}^*} h_{l,i}^{(N)}) 
+  \mbox{Im}(\lambda_k^* \lambda_l) \mbox{Im}( h_{k,i}^{{(N)}^*} h_{l,i}^{(N)})    {\big ) } }  {  N   }   {\Bigg \}}^2 \nonumber \\
& = & {\Bigg \{ } \sum_{i=1}^N {\Big ( } \sum_{k=1}^M \frac {\vert \lambda_k \vert^2 \vert h_{k,i}^{(N)} \vert^2 } {N}  {\Big ) }^2 {\Bigg \} } 
+ 4 {\Bigg [}
\sum_{k_1=1}^M \sum_{k_2=1}^M \sum_{l_2=k_2 + 1}^M 
 {\Bigg (} \vert \lambda_{k_1} \vert^2 \mbox{Re}(\lambda_{k_2}^* \lambda_{l_2}) \frac{ \sum_{i=1}^N  \vert h_{k_1,i}^{(N)} \vert^2 \mbox{Re}( h_{k_2,i}^{{(N)}^*}
 h_{l_2,i}^{(N)})  }{ N^2  } \nonumber \\
& &  \hspace{80mm} +  \vert \lambda_{k_1} \vert^2 \mbox{Im}(\lambda_{k_2}^* \lambda_{l_2}) \frac{ \sum_{i=1}^N  \vert h_{k_1,i}^{(N)} \vert^2 \mbox{Im}(
h_{k_2,i}^{{(N)}^*} h_{l_2,i}^{(N)})  }{ N^2  }    {\Bigg )} {\Bigg ]} \nonumber \\
& & + 4 \sum_{k_1=1}^M \sum_{k_2=1}^M \sum_{l_1=k_1+1}^M \sum_{l_2=k_2+1}^M {\Bigg \{} 
\,\,  \mbox{Re}(\lambda_{k_1}^* \lambda_{l_1}) \mbox{Re}(\lambda_{k_2}^* \lambda_{l_2}) \frac{\sum_{i=1}^N \mbox{Re}( h_{k_1,i}^{{(N)}^*} h_{l_1,i}^{(N)}) \mbox{Re}(
h_{k_2,i}^{{(N)}^*} h_{l_2,i}^{(N)}) }{ N^2    } + \nonumber \\
& & \hspace{50mm} \mbox{Re}(\lambda_{k_1}^* \lambda_{l_1}) \mbox{Im}(\lambda_{k_2}^* \lambda_{l_2}) \frac{\sum_{i=1}^N \mbox{Re}( h_{k_1,i}^{{(N)}^*} h_{l_1,i}^{(N)})
 \mbox{Im}( h_{k_2,i}^{{(N)}^*} h_{l_2,i}^{(N)}) }{ N^2    } + \nonumber \\
& &  \hspace{50mm} \mbox{Im}(\lambda_{k_1}^* \lambda_{l_1}) \mbox{Re}(\lambda_{k_2}^* \lambda_{l_2}) \frac{\sum_{i=1}^N \mbox{Im}( h_{k_1,i}^{{(N)}^*} h_{l_1,i}^{(N)}) 
\mbox{Re}( h_{k_2,i}^{{(N)}^*} h_{l_2,i}^{(N)}) }{ N^2   } + \nonumber \\
& &  \hspace{50mm} \mbox{Im}(\lambda_{k_1}^* \lambda_{l_1}) \mbox{Im}(\lambda_{k_2}^* \lambda_{l_2}) \frac{\sum_{i=1}^N \mbox{Im}( h_{k_1,i}^{{(N)}^*} h_{l_1,i}^{(N)}) 
\mbox{Im}( h_{k_2,i}^{{(N)}^*} h_{l_2,i}^{(N)}) }{ N^2    } {\Bigg \}}. 
\end{eqnarray}
}
\normalsize
From (As.2) in (\ref{assumptions}) it follows that for all $k_1,k_2,l_1,l_2 \in (1,2,\ldots,M)$
{
\small
\begin{eqnarray}
\label{step4_th3}
\lim_{N \rightarrow  \infty} \frac{\sum_{i=1}^N \mbox{Re}( h_{k_1,i}^{{(N)}^*} h_{l_1,i}^{(N)}) \mbox{Re}(
h_{k_2,i}^{{(N)}^*} h_{l_2,i}^{(N)}) }{ N^2    } = 0 \,,\,
\lim_{N \rightarrow  \infty} \frac{\sum_{i=1}^N \mbox{Re}( h_{k_1,i}^{{(N)}^*} h_{l_1,i}^{(N)})
 \mbox{Im}( h_{k_2,i}^{{(N)}^*} h_{l_2,i}^{(N)}) }{ N^2    } = 0 \nonumber \\
\lim_{N \rightarrow  \infty}  \frac{\sum_{i=1}^N \mbox{Im}( h_{k_1,i}^{{(N)}^*} h_{l_1,i}^{(N)}) 
\mbox{Re}( h_{k_2,i}^{{(N)}^*} h_{l_2,i}^{(N)}) }{ N^2   } = 0 \,,\,
\lim_{N \rightarrow  \infty} \frac{\sum_{i=1}^N \mbox{Im}( h_{k_1,i}^{{(N)}^*} h_{l_1,i}^{(N)}) 
\mbox{Im}( h_{k_2,i}^{{(N)}^*} h_{l_2,i}^{(N)}) }{ N^2    }  = 0 \nonumber \\
\lim_{N \rightarrow  \infty} \frac{ \sum_{i=1}^N  \vert h_{k_1,i}^{(N)} \vert^2 \mbox{Re}( h_{k_2,i}^{{(N)}^*}
 h_{l_2,i}^{(N)})  }{ N^2  } = 0 \,,\, \lim_{N \rightarrow  \infty}   \frac{ \sum_{i=1}^N  \vert h_{k_1,i}^{(N)} \vert^2 \mbox{Im}(
h_{k_2,i}^{{(N)}^*} h_{l_2,i}^{(N)})  }{ N^2  } = 0.
\end{eqnarray}
}
\normalsize
Substituting (\ref{step4_th3}) into (\ref{step3_th3}) and taking the limit, we have
{
\small
\begin{eqnarray}
\label{step6_th3}
\lim_{N \rightarrow \infty} \frac{8}{3} B_N^4 & = & \lim_{N \rightarrow \infty} {\Bigg \{ } \sum_{i=1}^N {\Big ( } \sum_{k=1}^M \frac {\vert \lambda_k \vert^2 \vert
h_{k,i}^{(N)} \vert^2 } {N}  {\Big ) }^2 {\Bigg \} }
\end{eqnarray}
}
\normalsize
Further,
{
\small
\begin{eqnarray}
\label{step7_th3}
\lim_{N \rightarrow \infty} {\Bigg \{ } \sum_{i=1}^N {\Big ( } \sum_{k=1}^M \frac {\vert \lambda_k \vert^2 \vert
h_{k,i}^{(N)} \vert^2 } {N}  {\Big ) }^2 {\Bigg \} } = 
\sum_{k_1 = 1}^M \sum_{k_2 = 1}^M {\Bigg (}  \vert \lambda_{k_1} \vert^2 \vert \lambda_{k_2} \vert^2  \lim_{N \rightarrow \infty} {\Big (} \frac{\sum_{i=1}^N \vert
 h_{k_1,i}^{(N)} \vert^2  \vert h_{k_2,i}^{(N)} \vert^2  }{ N^2 }  {\Big )}  {\Bigg )}
\end{eqnarray}
}
\normalsize
From (As.2) in (\ref{assumptions}) it follows that
%\begin{eqnarray}
%\label{step8_th3}
$\lim_{N \rightarrow \infty} {\Big (} \frac{\sum_{i=1}^N \vert
 h_{k_1,i}^{(N)} \vert^2  \vert h_{k_2,i}^{(N)} \vert^2  }{ N^2 }  {\Big )} = 0$
%\end{eqnarray}
and therefore using this result in (\ref{step7_th3}) and (\ref{step6_th3}) we get (\ref{stmt2_th3}).
%\begin{eqnarray}
%\label{step9_th3}
%\lim_{N \rightarrow \infty}   B_N^4 & = & 0.
%\end{eqnarray}
From (\ref{stmt1_th3}) it follows that $C_N^4$ converges to a positive constant as $N \rightarrow \infty$.
Hence we have now shown (\ref{th3_eq3_equiv}), and therefore the Lyapunov-CLT conditions
for the convergence of the c.d.f. of the random variable $\zeta_N$ are indeed satisfied.

Therefore invoking Result \ref{lyapunov_th} (Lyapunov-CLT), it follows that
the c.d.f. of $\zeta_N/C_N$ converges to the c.d.f. of a zero mean real Gaussian random variable with
unit variance.
Further, since $C_N$ converges to the constant $\sqrt{ \frac{1}{2} \sum_{k=1}^M c_k {\big (} (\lambda_k^I)^{2} + (\lambda_k^Q)^{2} {\big )} }$ (see (\ref{stmt1_th3})), using Result \ref{slutsky} (Slutsky's Theorem)
it follows that the c.d.f. of $\zeta_N$ converges to the c.d.f. of a zero mean real Gaussian random variable
with variance $\frac{1}{2} \sum_{k=1}^M c_k {\big (} (\lambda_k^I)^{2} + (\lambda_k^Q)^{2} {\big )}$.
$\hfill\blacksquare$
{
\vspace{-6mm}
}
\section{Probability of the Box Event ${\Big \{} {\bf z}_N \in {\mathcal B}_{_{\Delta}}({\bf u})  {\Big \}}$}
\label{box_event_proof}
\begin{mytheorem}\label{thm_continuity}
For a given channel sequence $ \{ {\bf H}_N \}_{N=M}^{\infty}$ satisfying (\ref{assumptions})
and a given fixed finite alphabet set ${\mathcal U}$, for any $\Delta > 0$, there exist a corresponding
integer $ N(\{ {\bf H}_N \},{\mathcal U}, \Delta)$, such that
for all $N \geq N(\{ {\bf H}_N \},{\mathcal U}, \Delta)$ (with fixed $M$)
{
\begin{eqnarray}
\mbox{Prob}({\bf z}_N \in {\mathcal B}_{_{\Delta}}({\bf u})) > 0 \,\,,\,\, \forall \, {\bf u} \in {\mathcal U}.
\end{eqnarray}
}
%\normalsize
where ${\mathcal B}_{_{\Delta}}({\bf u})$ is defined in (\ref{box_eqn}).
\end{mytheorem}

{\it Proof} --
To prove this result, we use the following expansion for the probability of a box event for any
general multivariate $n$-dimensional real r.v. ${\bf X} = (X_1,X_2,\cdots,X_n)$.
We consider the probability that ${\bf X}$ lies in a $n$-dimensional box centered at ${\boldsymbol \alpha} = (\alpha_1,\ldots,\alpha_n) \in {\mathbb R}^n$ and denoted by $C(\Delta,{\boldsymbol \alpha}) = {\big \{} (x_1, x_2, \cdots, x_n) \in {\mathbb R}^n \, \vert \, \alpha_k - \Delta \leq x_k \leq \alpha_k + \Delta
\,\,,\,\,k=1,2,\ldots,n{\big \}}$.
For notational convenience, we refer to $\alpha_k + \Delta$ and $\alpha_k - \Delta$
as the corresponding ``upper'' and ``lower'' limits for the $k$-th coordinate.
The probability that ${\bf X}$ lies in the box $C(\Delta,{\boldsymbol \alpha})$ is given by
{
\small
\begin{eqnarray}
\label{exp_box}
\mbox{Prob}({\bf X} \in C(\Delta,{\boldsymbol \alpha})) & = & \sum_{k=0}^n (-1)^k T_k(\Delta,{\boldsymbol \alpha})
\end{eqnarray}
}
\normalsize
where $T_k(\Delta,{\boldsymbol \alpha})$ is the probability that the r.v. $(X_1,X_2,\cdots,X_n)$ belongs to a sub-region of
${\big \{} (x_1,\cdots,x_n)  \in {\mathbb R}^n \,\, |\,\, x_l \leq \alpha_l + \Delta\,,\,l=1,2,\ldots,n{\big \}}$, where exactly $k$ coordinates are less
than their corresponding ``lower'' limit and the remaining $n - k$ coordinates are less than
their corresponding ``upper'' limit.
Specifically, $T_k(\Delta,{\boldsymbol \alpha})$ is given by\footnote{\footnotesize{
As an example, for $n=2$, we have
%\label{exp_box_exm}
$\mbox{Prob} {\Big ( } \alpha_1 - \Delta \leq X_1 \leq \alpha_1 + \Delta \,,\, \alpha_2 - \Delta \leq X_2 \leq \alpha_2 + \Delta {\Big ) }  =  T_0(\Delta,{\boldsymbol \alpha})  - T_1(\Delta,{\boldsymbol \alpha}) + T_2(\Delta,{\boldsymbol \alpha})$, where
%+ \mbox{Pr}(  X_1 \leq \alpha_1 + \Delta \,,\, X_2 \leq \alpha_2 + \Delta ) &  & \nonumber \\
%- \mbox{Pr}(  X_1 \leq \alpha_1 + \Delta \,,\, X_2 \leq \alpha_2 - \Delta ) &  & \nonumber \\
%- \mbox{Pr}(  X_1 \leq \alpha_1 - \Delta \,,\, X_2 \leq \alpha_2 + \Delta ) &  & \nonumber \\
%+ \mbox{Pr}(  X_1 \leq \alpha_1 - \Delta \,,\, X_2 \leq \alpha_2 - \Delta ) &  & \nonumber \\
$T_0(\Delta,{\boldsymbol \alpha})  \Define  \mbox{Prob}(  X_1 \leq \alpha_1 + \Delta \,,\, X_2 \leq \alpha_2 + \Delta )$, $T_2(\Delta,{\boldsymbol \alpha})  \Define  \mbox{Prob}(  X_1 \leq \alpha_1 - \Delta \,,\, X_2 \leq \alpha_2 - \Delta ) $, and
$T_1(\Delta,{\boldsymbol \alpha})  \Define \mbox{Prob}( X_1 \leq \alpha_1 + \Delta \,,\, X_2 \leq \alpha_2 - \Delta )
\,  + \, \mbox{Prob}(  X_1 \leq \alpha_1 - \Delta \,,\, X_2 \leq \alpha_2 + \Delta ) $.
%\end{eqnarray}
}}
{
\small
\begin{eqnarray}
\label{exp_box_2}
 T_k(\Delta,{\boldsymbol \alpha})  \,  =  \,   \sum_{i_1=1}^n\sum_{i_2=i_1+1}^n \cdots \sum_{i_k=i_{k-1}+1}^n 
\mbox{Prob}{\Big (} X_{r} \leq \alpha_{r} -\Delta\,\,\,\forall r \in \{i_1,i_2,\cdots,i_k \} 
\,\,,\,\,  X_{r} \leq \alpha_{r} + \Delta\,\,\,\forall r \notin \{i_1,i_2,\cdots,i_k \} {\Big )} 
\end{eqnarray}
}
\normalsize
Using the expansion in (\ref{exp_box}), the probability of the box event ${\Big \{} {\bf z}_N \in {\mathcal B}_{_{\Delta}}({\bf u})  {\Big \}}$ can be expressed as
{
\small
\begin{eqnarray}
\label{expansion_box}
\mbox{Prob}{\Big (} {\bf z}_N \in {\mathcal B}_{_{\Delta}}({\bf u})  {\Big )} =
\mbox{Prob} {\Big (} (\sqrt{E_k} u_k^I -  \Delta) \leq {z_k^I}^{(N)} \leq (\sqrt{E_k} u_k^I + \Delta) \,\,,\,\, 
(\sqrt{E_k} u_k^Q -  \Delta) \leq {z_k^Q}^{(N)} \leq (\sqrt{E_k} u_k^Q + \Delta) \nonumber \\
\hspace{70mm}  \,\,, \,\, k=1,2,\ldots,M {\Big )} \nonumber \\
\hspace{40mm}  =  \sum_{k=0}^{2M} (-1)^k \sum_{i_1=1}^{2M}\sum_{i_2=i_1+1}^{2M} \cdots \sum_{i_k=i_{k-1}+1}^{2M}
\mbox{Prob}{\Big (} z_{l}^{(N)} \leq \sqrt{E_l} u_{l} - \Delta\,\,\,\forall l \in \{i_1,i_2,\cdots,i_k \}  \,\,,\,\, \nonumber \\
\hspace{40mm} z_{l}^{(N)}  \leq  \sqrt{E_l} u_{l} + \Delta\,\,\,\forall l \notin \{i_1,i_2,\cdots,i_k \} {\Big )} 
\end{eqnarray}
}
\normalsize
where $z_l^{(N)}$ is the $l$-th component of ${\bf z}_N$ (i.e., $z_l^{(N)} = {z}^{Q^{(N)}}_{l/2}$ for even $l$, and $z_l^{(N)} =  z^{I^{(N)}}_{(l+1)/2}$ for odd
$l$) and $u_l$ is the $l$-th component
of the vector $(u_1^I,u_1^Q, u_2^I, u_2^Q, \cdots, u_M^I,u_M^Q)^T$.
For notational convenience we define
{
\small
\begin{eqnarray}
\label{TN_def}
T^{^{(N)}}(k,i_1,i_2,\cdots,i_k,{\bf u},\Delta) & \hspace{-2mm} \Define &
\hspace{-3mm} \mbox{Prob}{\Big (} z_{l}^{(N)} \leq \sqrt{E_l} u_{l} -\Delta\,\,\,\forall l \in \{i_1,i_2,\cdots,i_k \} 
\,\,,\,\,  z_{l}^{(N)}  \leq  \sqrt{E_l} u_{l} + \Delta\,\,\,\forall l \notin \{i_1,i_2,\cdots,i_k \} {\Big )}
 \nonumber \\ &  &  
\,\,\,\,\,\,\,1 \leq i_1 < i_2 < \cdots < i_k \leq 2M \,\,,\,\, 0 \leq k \leq 2M.
%\hspace{10mm} 1 \leq i_1 < i_2 < \cdots < i_k \leq 2M \,\,,\,\, 0 \leq k \leq 2M. &
\end{eqnarray}
}
\normalsize
Let ${\bf Y} = (Y_1, Y_2, \cdots, Y_{2M})$ denote a multivariate $2M$-dimensional real Gaussian random variable with independent zero mean components and
$\mbox{var}(Y_{2k-1}) = \mbox{var}(Y_{2k}) = c_k/2 \,,\,k=1,2,\ldots,M$.
%Also, let us define
%\begin{eqnarray}
%\label{Y_box}
%{\mathcal E}_Y({\bf u},\Delta) & = & \{   u_l - \Delta \leq Y_l \leq u_l + \Delta\,\,,\,\,l=1,2,\cdots,2M \} \nonumber \\
%\end{eqnarray}
From Theorem \ref{thm_conv_zN} it follows that the c.d.f. of ${\bf z}_N$ converges to the c.d.f. of ${\bf Y}$ in the limit as $N \rightarrow \infty$.
This convergence in distribution implies that, for any given arbitrary $\delta > 0$, for each term $T^{^{(N)}}(k,i_1,i_2,\cdots,i_k,{\bf u},\Delta)$, there
exists a corresponding positive integer $N(k,i_1,i_2,\cdots,i_k,\delta,{\bf u},\Delta)$ such that
for all $ N \geq N(k,i_1,i_2,\cdots,i_k,\delta,{\bf u},\Delta)$
{
\small
\begin{eqnarray}
\label{delta_nbh}
{\Big \vert}
T^{^{(N)}}(k,i_1,i_2,\cdots,i_k,{\bf u},\Delta) & - &
\mbox{Prob}{\Big (} Y_{l} \leq \sqrt{E_l} u_{l} - \Delta\,\,\,\forall l \in \{i_1,i_2,\cdots,i_k \}  
\,\,,\,\, \nonumber \\ & &  \hspace{12mm} Y_{l}  \leq \sqrt{E_l} u_{l} + \Delta\,\,\,\forall l \notin \{i_1,i_2,\cdots,i_k \} {\Big )}\,  {\Big \vert} \,\, \leq \,\, \delta.
\end{eqnarray}
}
\normalsize
We then choose a positive integer $g{\big (}\{ {\bf H}_N \},{\bf u},\Delta,\delta{\big )}$ given by
{
\small
\begin{eqnarray}
\label{chooseN_deltau}
g{\big (} \{ {\bf H}_N \}, {\bf u},\Delta,\delta {\big )} & \Define   & 
 \,\, \max_{k=0,1,\cdots,2M}\,\,\, \max_{1 \leq i_1 < i_2 < \cdots < i_k \leq 2M} \,\,
  N(k,i_1,i_2,\cdots,i_k, \delta,{\bf u},\Delta) 
\end{eqnarray}
}
\normalsize
%It is now clear that, for any $N \geq g(\delta,{\bf u},\Delta)$, the absolute difference
%between $T(k,i_1,i_2,\cdots,i_k,\delta,{\bf u},\Delta)$ and the corresponding probability
%with the limiting distribution (i.e., distribution of $Y$), is less than
%$\delta$ for any $k \in (0,1,\cdots,2M)$ and for each $k$
%any choice of increasing integers $1 \geq i_1 < i_2 < \cdots < i_k \leq 2M$.
Combining (\ref{expansion_box}), (\ref{TN_def}) and (\ref{delta_nbh}), for all $N \geq g{\big (} \{ {\bf H}_N \},{\bf u},\Delta,\delta{\big )}$ we have
{
\small
\begin{eqnarray}
\label{expansion_box_final}
& \hspace{-290mm} {\Big \vert} \mbox{Prob}{\big (}{\bf z}_N \in {\mathcal B}_{_{\Delta}}({\bf u})  {\big )} -  \mbox{Prob}{\big (}{\bf Y} \in {\mathcal B}_{_{\Delta}}({\bf u})  {\big )} {\Big \vert} &  \nonumber \\
 = \,\, {\Bigg \vert} \sum_{k=0}^{2M} \sum_{i_1=1}^{2M}\sum_{i_2=i_1+1}^{2M} \hspace{-2mm} ... \sum_{i_k=i_{k-1}+1}^{2M} \hspace{-4mm} (-1)^k {\Bigg \{} 
T^{^{(N)}}(k,i_1,i_2,\cdots,i_k,{\bf u},\Delta) \,\,  - \,\, \mbox{Prob}{\Big (} Y_{l} \leq \sqrt{E_l} u_{l} -\Delta\,\,\,\forall l \in \{i_1,i_2,\cdots,i_k \} \,\,,\,\, & & 
\nonumber \\
&  \hspace{-78mm} Y_{l}  \leq \sqrt{E_l} u_{l} + \Delta\,\,\,\forall l \notin \{i_1,i_2,\cdots,i_k \} {\Big )}    {\Bigg \}} {\Bigg \vert} & \nonumber \\
 \leq \,\, \sum_{k=0}^{2M} \sum_{i_1=1}^{2M}\sum_{i_2=i_1+1}^{2M}  ...  \sum_{i_k=i_{k-1}+1}^{2M} {\Bigg \vert}  {\Bigg \{}
T^{^{(N)}}(k,i_1,i_2,\cdots,i_k,{\bf u},\Delta) \, - \, \mbox{Prob}{\Big (} Y_{l} \leq \sqrt{E_l}u_{l} -\Delta\,\,\,\forall l \in \{i_1,i_2,\cdots,i_k \} \,\,,\,\, & &
\nonumber \\
&  \hspace{-78mm} Y_{l}  \leq \sqrt{E_l} u_{l} + \Delta\,\,\,\forall l \notin \{i_1,i_2,\cdots,i_k \} {\Big )}    {\Bigg \}} {\Bigg \vert} & \nonumber \\
& \hspace{-260mm}  \leq \,\,  \sum_{k=0}^{2M} \sum_{i_1=1}^{2M}\sum_{i_2=i_1+1}^{2M} \cdots \sum_{i_k=i_{k-1}+1}^{2M} \delta \,
\,  = \,\,    2^{2M} \delta. &
\end{eqnarray}
}
\normalsize
Since the range space (support) of ${\bf Y}$ is the entire space ${\mathbb R}^{2M}$, it 
follows that $\mbox{Prob}{\big (}{\bf Y} \in {\mathcal B}_{_{\Delta}}({\bf u})  {\big )} > 0$ (i.e., strictly positive)
for any $\Delta > 0$ and all ${\bf u} \in {\mathcal U}$.
For the given information symbol vector ${\bf u}$ and $\Delta > 0$, we choose a corresponding $\delta$ given by
{
\small
\begin{eqnarray}
\label{delta_star}
\delta({\bf u},\Delta) \Define \frac{1}{2}  \frac{\mbox{Prob}{\big (}{\bf Y} \in {\mathcal B}_{_{\Delta}}({\bf u})  {\big )}} { 2^{2M} } \,\, > \,\, 0
\end{eqnarray}
}
\normalsize
%where $\rho$ is an arbitrarily chosen number in the open interval ${\big (} 0 \,,\, \frac{\mbox{Pr}({\mathcal E}_Y({\bf u},\Delta))} { 2^{2M} }{\big )}$.
From (\ref{expansion_box_final}) and (\ref{delta_star}) it now follows that, for all $N > g{\big (} \{ {\bf H}_N \},{\bf u},\Delta,\delta({\bf u},\Delta){\big )}$ we have
{
\small
\begin{equation}
{\Big \vert} \mbox{Prob}{\big (}{\bf z}_N \in {\mathcal B}_{_{\Delta}}({\bf u})  {\big )} -  \mbox{Prob}{\big (}{\bf Y} \in {\mathcal B}_{_{\Delta}}({\bf u})  {\big )} {\Big \vert} \leq 2^{2M} \delta({\bf u},\Delta) = \frac {\mbox{Prob}{\big (}{\bf Y} \in {\mathcal B}_{_{\Delta}}({\bf u})  {\big )}} {2}
\end{equation}
}
\normalsize
which then implies that
{
\small
\begin{equation}
\label{delta_star1}
\mbox{Prob}{\big (}{\bf z}_N \in {\mathcal B}_{_{\Delta}}({\bf u})  {\big )} \geq \frac{\mbox{Prob}{\big (}{\bf Y} \in {\mathcal B}_{_{\Delta}}({\bf u})  {\big )}} {2} > 0
\end{equation}
}
\normalsize
i.e., $\mbox{Prob}{\big (}{\bf z}_N \in {\mathcal B}_{_{\Delta}}({\bf u})  {\big )} $ is strictly positive for $N > g{\big (} \{ {\bf H}_N \},{\bf u},\Delta,\delta({\bf u},\Delta){\big )}$.
For a given channel sequence $\{ {\bf H}_N \}$, a finite ${\mathcal U}$ and $\Delta > 0$, we define the integer
{
\small
\begin{eqnarray}
\label{def_N_HUD}
N(\{ {\bf H}_N \},{\mathcal U}, \Delta) & \Define & \max_{u \in {\mathcal U}} g{\big (} \{ {\bf H}_N \},{\bf u},\Delta,\delta({\bf u},\Delta){\big )}.
\end{eqnarray}
}
\normalsize
Combining this definition with the result in (\ref{delta_star1}) proves the theorem.
$\hfill\blacksquare$

\newpage

\begin{table}
\caption{Minimum $P_T/\sigma^2$ (DB) required to achieve a per-user ergodic rate of $2$ bpcu}
\centering
\vspace{-4mm}
\begin{tabular}{| c ||  c |  c | c | c | c | c | c | c | c |}
\hline
 & N=60 & N=80 & N=100 & N=120 & N=160 & N=200 & N=240 & N = 320 & N = 400   \\
\hline
GBC Sum Capacity Upper Bound (M = 10) & -2.8 & -4.0 & -5.1 & -5.8 & -7.2 & -8.2 & -8.9 & -10.2 & -11.2 \\
\hline
Proposed CE Precoder (M = 10) & -0.8 & -2.1 & -3.3 & -4.1 & -5.5 & -6.5 & -7.2 & -8.6 & -9.6 \\
\hline
Power Gap (M = 10) & 2.0 & 1.9 & 1.8 & 1.7 & 1.7 & 1.7 & 1.7 & 1.6 & 1.6 \\
\hline
GBC Sum Capacity Upper Bound (M = 40) & 3.8 & 2.4 & 1.3  & 0.6 & -0.9 & -2.0 & -2.7 & -4.1 & -5.1 \\
\hline
Proposed CE Precoder (M = 40)  & 9.2 & 6.0 & 4.1  & 3.2 & 1.4 & -0.1 & -0.9 & -2.3 & -3.5 \\
\hline
Power Gap (M = 40) & 5.4 & 3.6 & 2.8 & 2.6 & 2.3 & 1.9 & 1.8 & 1.8 & 1.6 \\
\hline
%Power Gap  & 4.2 & 4.0 & 4.7  & 5.3 & 5.4 & 5.3 \\
%\hline
\end{tabular}
\vspace{-6mm}
\label{table_1}
\end{table}

{\vspace{-3mm}}


\end{document}